\documentclass[iop]{emulateapj}
\usepackage{graphics}
\usepackage{amsmath}
\usepackage{empheq} 
\usepackage{rotating}
\newcommand{\degree}{\hbox{$^\circ$}}

\newcommand{\ltsimeq}{\la}
\newcommand{\gtsimeq}{\ga}

\newcommand{\msun}{M$_{\odot}$}

\newcommand{\HI}{H{\sc i}}

\renewcommand*{\thefootnote}{\fnsymbol{footnote}} 

\shortauthors{McQuinn et al.}
\shorttitle{Distances to SHIELD Galaxies}

\begin{document}
\title{Distance Determinations to SHIELD Galaxies from HST Imaging\protect\footnotemark[*]}
\footnotetext[*]{Based on observations made with the NASA/ESA Hubble Space Telescope, obtained from the Data Archive at the Space Telescope Science Institute, which is operated by the Association of Universities for Research in Astronomy, Inc., under NASA contract NAS 5-26555.}
\author{Kristen.~B.~W. McQuinn\altaffilmark{1}, 
John M.~Cannon\altaffilmark{2},
Andrew E.~Dolphin\altaffilmark{3},
Evan D. Skillman\altaffilmark{1},
John J.~Salzer\altaffilmark{4},
Martha P.~Haynes\altaffilmark{5},
Elizabeth Adams\altaffilmark{5},
Ian Cave\altaffilmark{2}, 
Ed C. Elson\altaffilmark{6},
Riccardo Giovanelli\altaffilmark{5},
Ju\"{e}rgen Ott\altaffilmark{7},
Am\'elie Saintonge\altaffilmark{8}
}

\altaffiltext{1}{Minnesota Institute for Astrophysics, School of Physics and
Astronomy, 116 Church Street, S.E., University of Minnesota,
Minneapolis, MN 55455, \ {\it kmcquinn@astro.umn.edu}} 
\altaffiltext{2}{Department of Physics and Astronomy, 
Macalester College, 1600 Grand Avenue, Saint Paul, MN 55105}
\altaffiltext{3}{Raytheon Company, 1151 E. Hermans Road, Tucson, AZ 85756}
\altaffiltext{4}{Department of Astronomy, Indiana University, 727 East 3rd Street, Bloomington, IN 47405}
\altaffiltext{5}{Center for Radiophysics and Space Research, Space Sciences Building, Cornell University, Ithaca, NY 14853}
\altaffiltext{6}{Astrophysics, Cosmology and Gravity Centre (ACGC), Department of Astronomy, University of Cape Town, Private Bag X3, Rondebosch 7701, South Africa}
\altaffiltext{7}{National Radio Astronomy Observatory, P.O. Box O, 1003 Lopezville Road, Socorro, NM 87801, USA}
\altaffiltext{8}{Department of Physics and Astronomy, University College London, London WC1E 6BT, UK}

\begin{abstract}
The Survey of \HI\ in Extremely Low-mass Dwarf galaxies (SHIELD) is an on-going multi-wavelength program to characterize the gas, star formation, and evolution in gas-rich, very low-mass galaxies. The galaxies were selected from the first $\sim10$\% of the \HI\ ALFALFA survey based on their inferred low \HI\ mass and low baryonic mass, and all systems have recent star formation. Thus, the SHIELD sample probes the faint end of the galaxy luminosity function for star-forming galaxies. Here, we measure the distances to the 12 SHIELD galaxies to be between $5-12$ Mpc by applying the tip of the red giant method to the resolved stellar populations imaged by the Hubble Space Telescope. Based on these distances, the \HI\ masses in the sample range from $4\times10^6$ to $6\times10^7$ \msun, with a median \HI\ mass of $1\times 10^7$ \msun. The TRGB distances are up to 73\% farther than flow-model estimates in the ALFALFA catalog. Because of the relatively large uncertainties of flow model distances, we are biased towards selecting galaxies from the ALFALFA catalog where the flow model underestimates the true distances. The measured distances allow for an assessment of the native environments around the sample members. Five of the galaxies are part of the NGC~672 and NGC~784 groups, which together constitute a single structure. One galaxy is part of a larger linear ensemble of 9 systems that stretches 1.6 Mpc from end to end. Two galaxies reside in regions with $1-4$ neighbors, and four galaxies are truly isolated with no known system identified within a radius of 1 Mpc. 
\end{abstract} 


\section{Introduction\label{intro}}
Historically, optical surveys have cataloged the majority of dwarf galaxies in the nearby universe. However, optical identification of systems populating the extremely low-mass end of the galaxy luminosity function are hampered by the intrinsic faintness and small angular size of such systems. These low luminosity systems dominate the local population of galaxies and are important tracers of the distribution of mass. Outside of a dense group or cluster environment where low mass galaxies do not repeatedly experience ram pressure and/or tidal stripping of their ISM, populations of such extremely low mass dwarfs are expected to be predominantly gas-rich. Thus, large \HI\ surveys offer an opportunity to identify and catalogue such systems in a statistically complete volume.

The Arecibo Legacy Fast ALFA (ALFALFA) survey \citep{Giovanelli2005, Haynes2011} is a blind extragalactic \HI\ survey to map the nearby \HI\ universe over 7000 sq. degrees of high Galactic latitude sky. With an \HI\ mass detection limit as low as $10^6$ \msun\ for galaxies in the local universe and $10^{9.5}$ \msun\ at the survey velocity limit of $z\sim0.06$, the ALFALFA survey was designed to populate the faint end of the \HI\ mass function over a cosmologically significant volume. The low mass end of the \HI\ mass function has been sampled by previous studies including the wide area \HI\ Parkes All-Sky Survey \citep[HIPASS;][]{Barnes2001, Meyer2004, Wong2006}. However, ALFALFA has higher angular and spectral resolution and higher sensitivity. Thus, the ALFALFA catalogue provides a representative sampling of the \HI\ mass function to lower masses in a much larger volume. 

The Survey of \HI\ in Extremely Low-mass Dwarf systems \citep[SHIELD;][]{Cannon2011} was designed to probe the early release of the ALFALFA dataset in a systematic investigation of nearby galaxies with HI masses $\ltsimeq 10^7$ \msun\ outside the Local Group. From the first $\sim10$\% of the processed ALFALFA survey data, 12 systems were selected that had gas mass estimates between $10^6-10^7$ \msun\ based on \HI\ line widths and flow-model distances, and overall low masses based on an \HI\ full width at half maximum $< 65$ km s$^{-1}$ which discriminated against gas-poor massive galaxies. The high positional accuracy of the ALFALFA survey (i.e., better than 20$\arcsec$) facilitated the identification of optical counterparts in SDSS imaging, revealing young, blue stellar populations in each galaxy. These low mass systems have retained \HI\ mass reservoirs of $10^6-10^7$ \msun\ over a Hubble time, apparently evolving in relative quiescence, and all have recent star formation activity, consistent with all previous observations of dwarf irregular galaxies \citep[e.g.,][]{Hunter1982}. 

Once the presence of stellar components was confirmed with SDSS imaging, follow-up observations became possible in order to obtain accurate distance measurements and to probe the evolutionary histories of the galaxies. Accurate distances to these very low mass galaxies are important for a number of reasons. First, accurate distances to galaxies reduce uncertainties in the \HI\ mass function made possible by the ALFALFA survey. Second, distance measurements enable distance dependent analyses of the individual systems to be placed on an absolute scale. For example, the distance measurement of 1.72$^{+0.14}_{-0.40}$ to a newly discovered, extremely metal-deficient galaxy from the ALFALFA survey, Leo~P, allows interpretation of the star formation and evolutionary processes to be placed on solid ground \citep{Giovanelli2013, Rhode2013, Skillman2013, McQuinn2013}. Third, combining \HI\ velocity measurements and accurate distances allows an assessment of environment. Most dwarf galaxies are thought to exist in some form of association with other galaxies \citep{Tully2006}, with truly isolated galaxies the exception even in low density environments. The Nearby Catalog of Galaxies \citep{Karachentsev2013} compiles measurements of 869 galaxies within 11 Mpc, including calculations of tidal indices that parameterize the likelihood of neighboring galaxies being kinematically bound in groups. Clearly, such work depends critically on accurate distance measurements. Finally, \HI\ velocity measurements and accurate distances enable low mass galaxies to be used to help trace the distribution of matter in larger low-density galaxy volumes. Significant effort has been made to map the distribution of galaxies locally. The very local Hubble flow ($0.7$ Mpc $<$ D$_{MW} < 3.0$ Mpc) has been mapped with increasing accuracy, measuring the expansion of the universe outside the boundaries of the Local Group \citep{Karachentsev2009}. The Extragalactic Distance Database \citep{Tully2009}, the Cosmicflow program \citep{Courtois2011a, Courtois2011b, Tully2012}, and the Cosmicflows$-$2 program \citep{Tully2013} are working to expand this analysis out to larger distances. The population of low mass galaxies with accurate distances from the richly populated ALFALFA data set can make an important contribution to mapping the mass in the Local Volume (LV; defined by an approximate radius of 11 Mpc). 

Here, we present the distance measurements to the 12 SHIELD galaxies based on Hubble Space Telescope (HST) optical imaging and the tip of the red giant branch (TRGB) standard candle method \citep[e.g.,][]{Mould1986}. The paper is organized as follows. \S2 describes the sample and summarizes the observations and data processing. \S3 describes the distance determination methods and results. \S4 maps the native environment around the SHIELD galaxies including all known galaxies within a radius of 1 Mpc of each system. \S5 summarizes our conclusions. Future work on the SHIELD sample will include an investigation of their star formation histories (McQuinn et~al., in preparation), metallicity measurements based on nebular abundance analyses (Haurberg et~al., in preparation), and a study of the gas kinematics from VLA observations (Cannon et~al., in preparation).

\section{The Galaxy Sample, Observations, and Photometry Method\label{obs}}
Table~\ref{tab:galaxies} lists the 12 galaxies that comprise the SHIELD sample, their coordinates and observation details. The observations were obtained with the HST using the Advanced Camera for Surveys (ACS) Wide Field Channel (WFC) \citep{Ford1998} and are a combination of new and archival data. Eleven galaxies were part of the HST-GO-12658 SHIELD program (PI: Cannon). Three galaxies were part of the previous HST-GO-10210 program (PI: Tully); the targets of two of these three archival data sets overlap with the SHIELD program. The observations were obtained with two filters during single HST orbits and were cosmic-ray split (CRSPLIT=2). Average integration times were 1000 s in the F606W filter and 1200 s in the F814W filter. The observations for the two galaxies that overlapped in the observing programs (AGC~111977; AGC~174585) have longer final integration times of $\sim$1900 s in the F606W filter and $\sim$2300 s in the F814W filter. The observing programs and final integration times for each galaxy are noted in Table~\ref{tab:galaxies}.

The images were cosmic-ray cleaned and processed by the standard ACS pipeline.  The raw images showed the effects of charge transfer efficiency (CTE) non-linearities caused by space radiation damage on the ACS instrument \citep[e.g.,][]{Anderson2010, Massey2010}. The CTE streaks in the images were cosmetically corrected using the HST procedure \textsc{pixctecorr} before being combined using \textsc{MultiDrizzle}.

\begin{figure*}[ht]
\epsscale{1.0}
\plotone{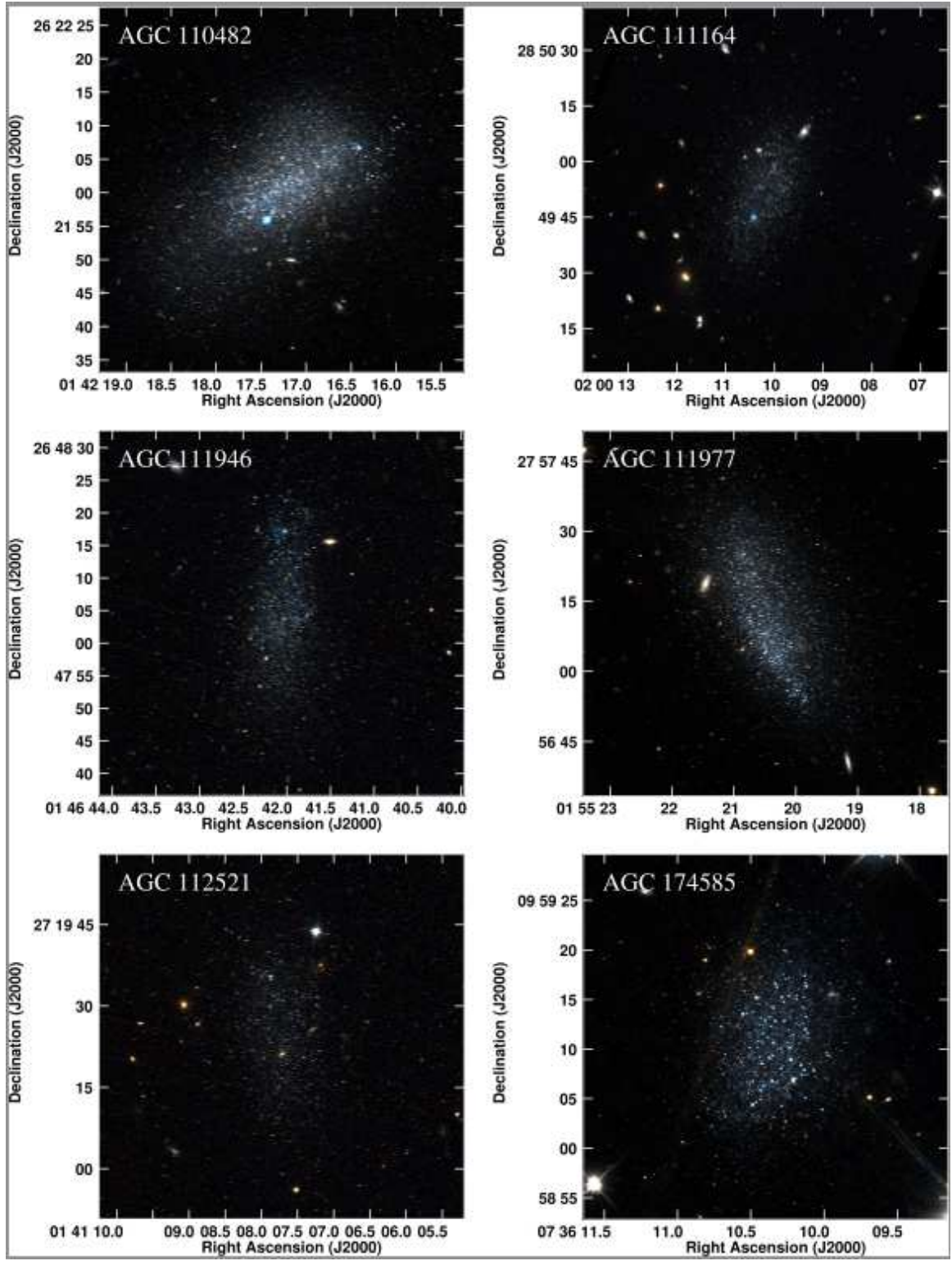}
\caption{HST 3-color images of the SHIELD sample combining F606W images (blue), the average of F606W and F814W images (green), and F814W images (red). North is up and East is left.  Each field of view encompasses twice the optical major axis as determined by iteratively plotting CMDs with different elliptical parameters (see below and Table~\ref{tab:galaxies}, Columns 9$-$11). The galaxies host young, blue stellar populations as well as older, red underlying stellar population. A high-resolution version of this image is available in the Astrophysical Journal.}
\label{fig:images}
\end{figure*}

\begin{figure*}[h]
\addtocounter{figure}{-1}
\epsscale{1.0}
\plotone{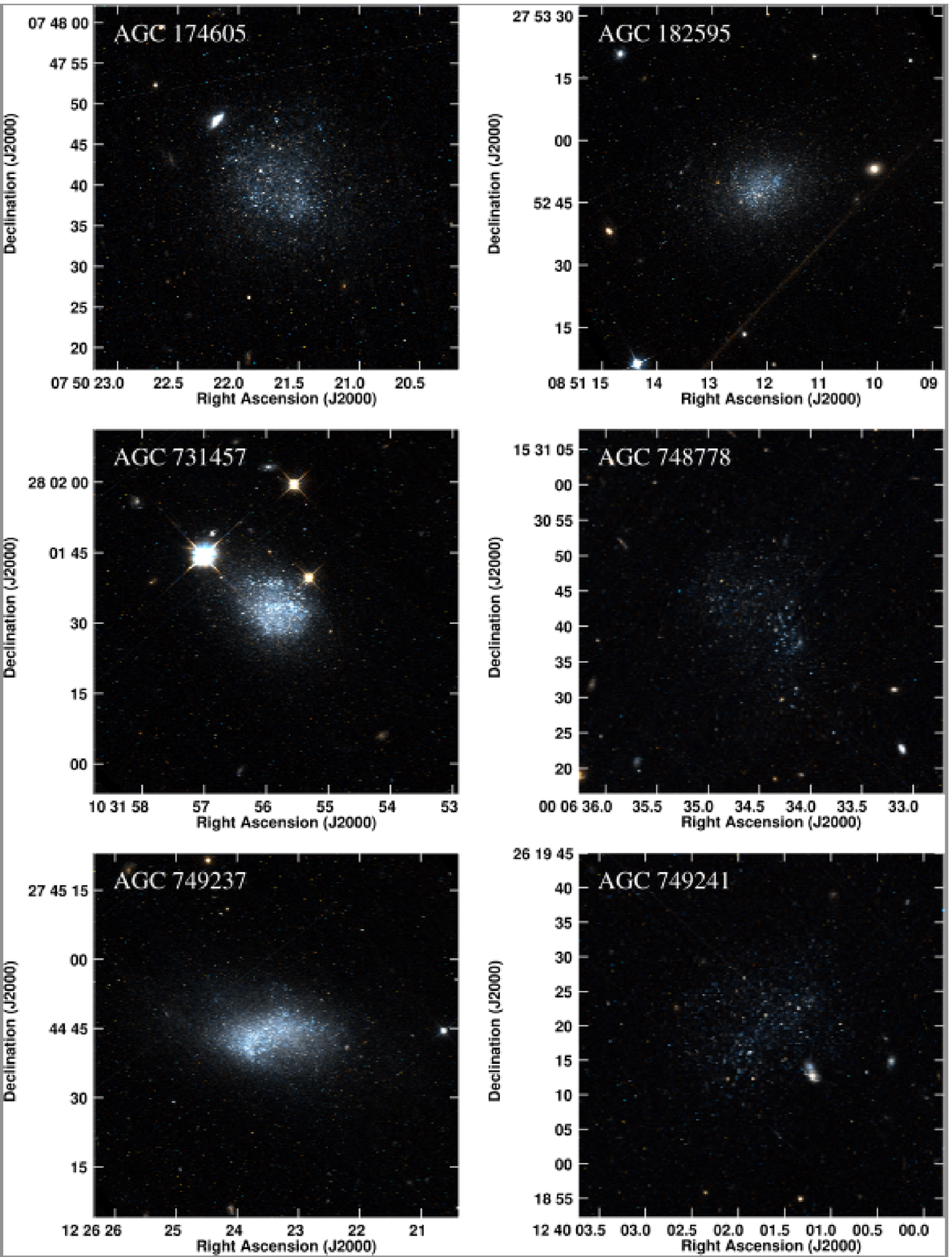}
\caption{\textit{Continued.}}
\end{figure*}

Figure~\ref{fig:images} shows the final 3-color multidrizzled images combining the F606W (blue), the average of the F606W and F814W images (green), and F814W (red) observations. The images are oriented with North up and East left. Each field of view encompasses twice the optical major axis as determined by iteratively plotting CMDs with different elliptical parameters (see below and Table~\ref{tab:galaxies}).  Common to each system is a concentrated young, blue stellar population and an extended underlying red stellar population. There are also differences in morphology ranging from the rather compact nature of AGC~182595, AGC~731457, and AGC~749237, to the more elongated distribution of stars in AGC~110482, AGC~111977, and AGC~174585. 

Photometry was performed on the pipeline processed, non-CTE corrected, cosmic-ray cleaned images (CRJ files) using the ACS module of the DOLPHOT photometry package \citep{Dolphin2000}. CTE corrections were performed using the CTE routine within DOLPHOT. The resulting photometry list was filtered on a number of the parameters measured for each point source. Only sources classified as stars with error flags $<$4 were kept. Cuts were made on the sharpness and crowding parameters. Sharpness indicates whether a point source is too broad (such as background galaxies) or too sharp (such as cosmic rays). Crowding measures how much brighter a star would be if nearby stars had not been fit simultaneously; stars with higher values of crowding have higher photometric uncertainties. Specifically, we rejected point sources with (V$_{sharp}+$I$_{sharp}$)$^2>0.075$ and (V$_{crowd}+$I$_{crowd}$)$>0.8$. 

The photometry was also filtered on the signal-to-noise ratios (SNR) of sources in each filter. The F814W photometry was filtered for sources with SNR$>5\sigma$, ensuring only sources with high fidelity measurements are used as input to the TRGB measurement. The F606W photometry was filtered twice. First, we made a cut on sources with a SNR$\geq5\sigma$ for the purposes of creating clean color-magnitude diagrams (CMDs). Second, we made a cut to include sources with a SNR$>2.5\sigma$. This more liberal cut in the F606W photometry avoids excluding sources with a low SNR in the F606W filter but higher SNRs in the F814W photometry, preventing the introduction of completeness effects when selecting sources for the TRGB measurements. Artificial star tests were performed to measure the completeness limit of the images using the same photometry package and filtered on the same parameters. The photometry lists were corrected for Galactic absorption based on the dust maps of \citet{Schlegel1998}; these values are noted in Table~\ref{tab:galaxies}.

Spatial cuts were applied to the photometry. These cuts were determined iteratively by plotting the color magnitude diagram (CMD) of stars from concentric ellipses centered on the optical center of each galaxy with ellipticities and position angles that approximately matched the stellar distributions. Using point sources from the inner ellipses, the CMDs are dominated by stars from the galaxies. In contrast, when using point sources from the outer ellipses, the CMDs are dominated by background sources. The semi-major and semi-minor axes of the ellipses were increased until the CMDs from larger annuli matched the distribution of point sources from a field region CMD. These empirically determined estimates of the semi-major axes, ellipticities, and position angle values are listed in Table~\ref{tab:galaxies}. 

Figure~\ref{fig:cmds} shows the CMDs from the photometry with SNR$>5\sigma$ in both filters. The depth of the photometry corresponds to the 50\% completeness level determined from the artificial star tests. Representative photometric uncertainties per magnitude are shown. The photometric depth of the CMDs varies indicating a range in the distances to the galaxies. Most importantly for distance measurements, each CMD has a detection of the red giant branch (RGB) sequence.

\begin{figure*}[ht]
\includegraphics[width=\textwidth]{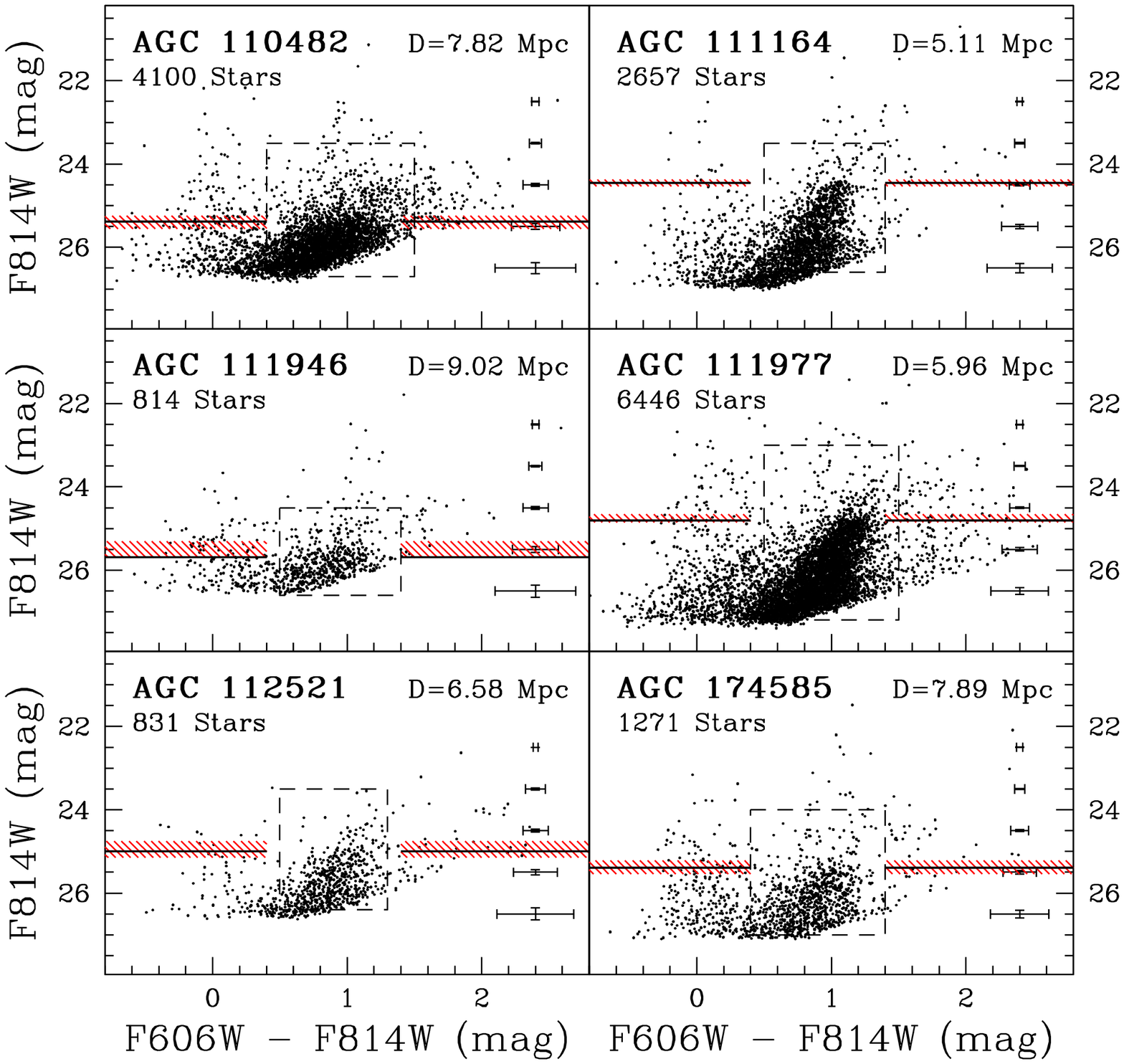}
\caption{CMDs of the SHIELD sample. Representative uncertainties per magnitude are shown at the right. In each CMD, the RGB is clearly detected, along with a population of young MS and HeB stars. The dashed black line outlines the region of RGB used in the TRGB measurement. The shaded red line marks the break in the F814W LF identified by the Sobel filter including uncertainties; the solid black line marks the luminosity of the TRGB identified by the ML technique. The final distance measurements are shown in the upper right corner of each CMD.}
\label{fig:cmds}
\end{figure*}

\begin{figure*}[ht]
\addtocounter{figure}{-1}
\includegraphics[width=\textwidth]{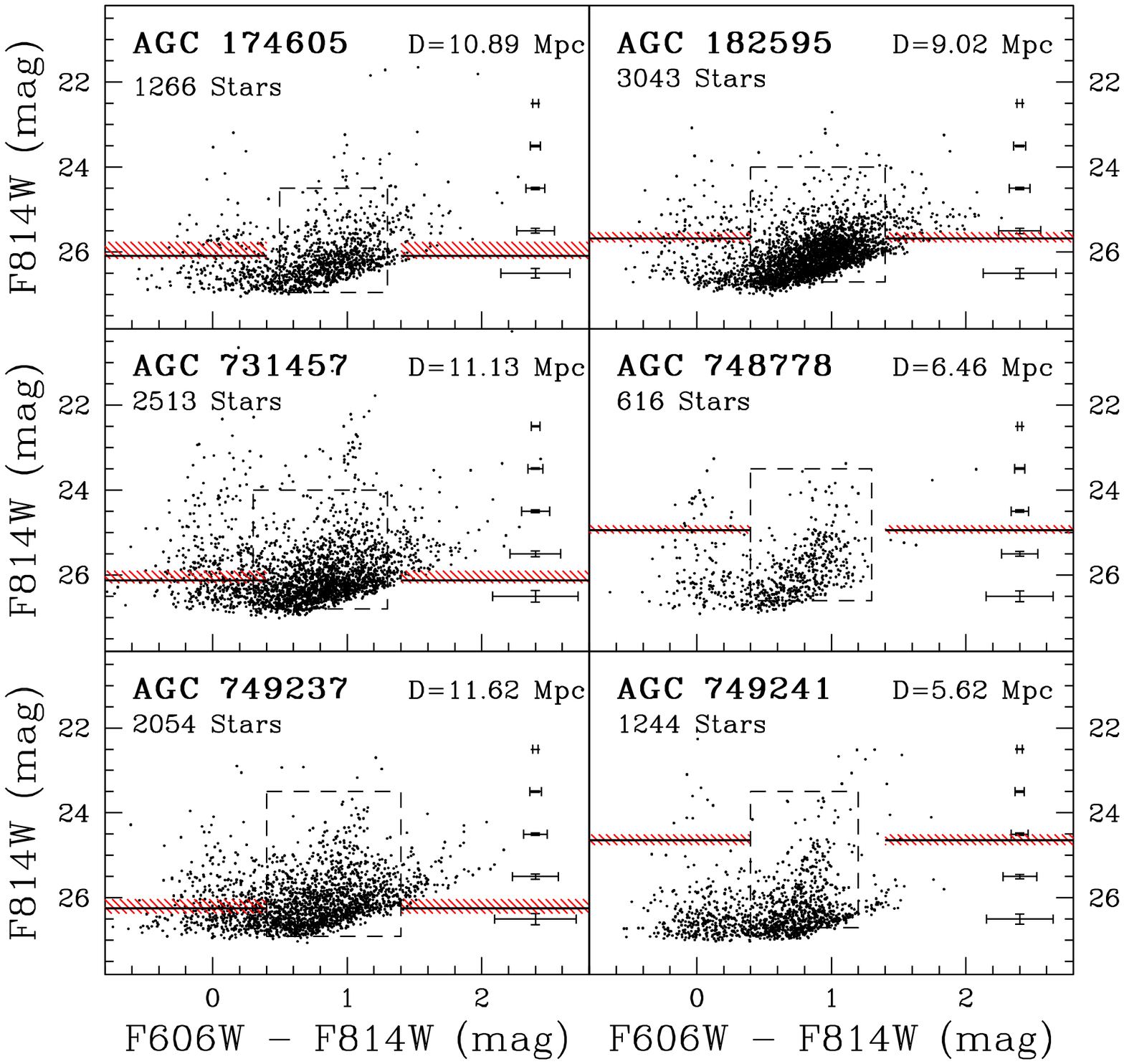}
\caption{\textit{Continued}}
\end{figure*}

\begin{figure}
\includegraphics[width=\columnwidth]{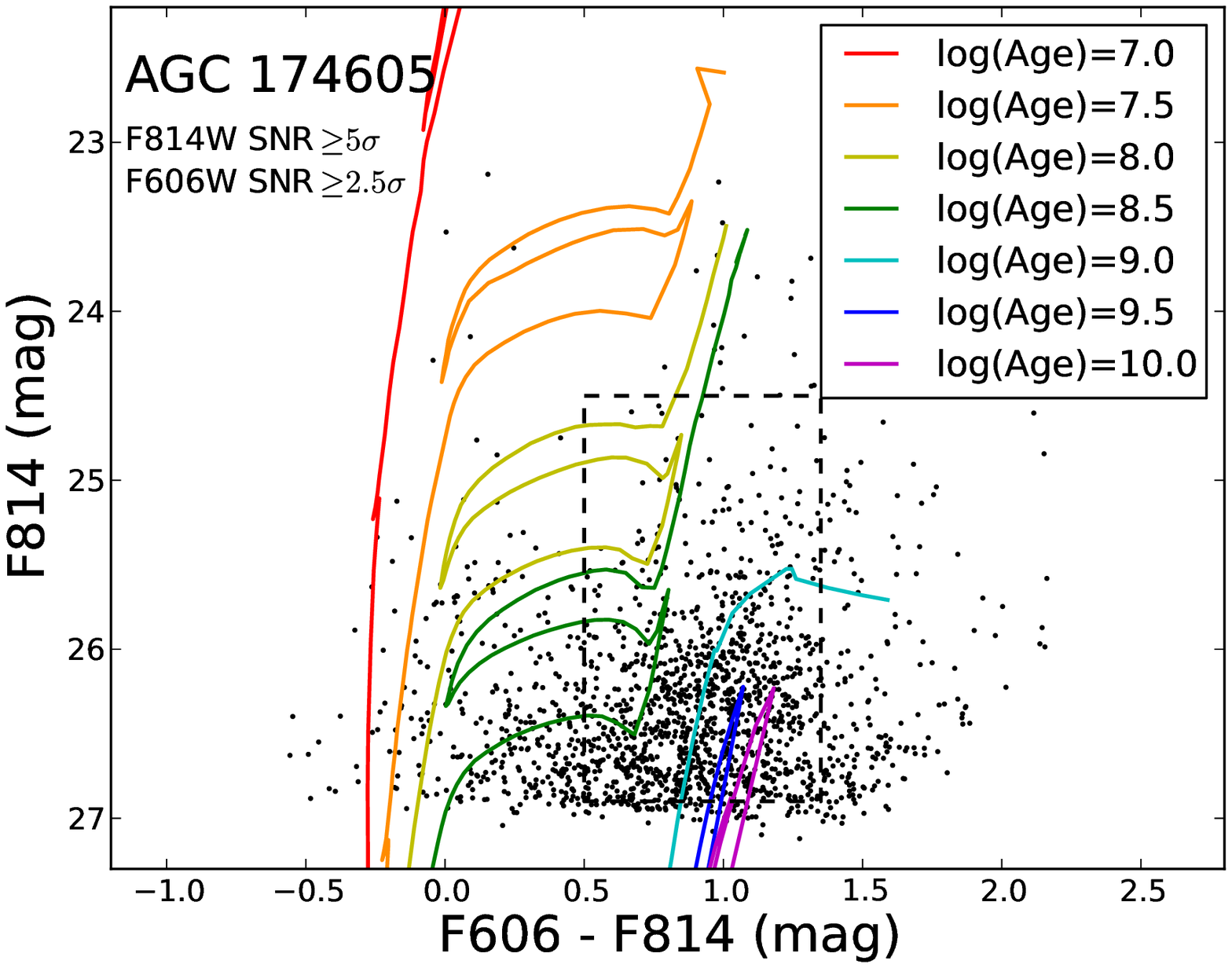}
\caption{CMD of AGC~174605 with the F606W photometry filtered with a more liberal SNR cut of 2.5$\sigma$, while the F814W photometry remains filtered with the higher SNR cut of 5$\sigma$. This more liberal SNR allowance in the F606W filter avoids introducing completeness effects in the F814W photometry used in the TRGB measurements. The dashed black line outlines the region of RGB stars used in the TRGB measurement. Padova isochrones (z$=$0.002) are overplotted in log(Age) from 7.0-10.0 in increments of 0.5 dex to show the approximate locations of stars of different ages and stages of stellar evolution.}
\label{fig:cmd_Vsnr2_5}
\end{figure}

Figure~\ref{fig:cmd_Vsnr2_5} re-plots the CMD for AGC~174605 with the F606W photometry using the more liberal SNR$\geq2.5\sigma$ and the same F814W photometry. Note the lower part of the CMD is more fully populated compared to Figure~\ref{fig:cmd_Vsnr2_5}, thus enabling a more complete selection of stars to be included in the F814W LF. Overlaid on the CMD are Padova isochrones \citep[z$=$0.002][]{Girardi2010} which highlight the expected locations of stars of different ages in the CMD. Similar to all of the CMDs shown in Figure~\ref{fig:cmds}, the upper main sequence (MS) is populated, indicating that the galaxies host sites of recent star formation. This is further confirmed by the presence of blue and red helium burning sequences \citep[cf.,][]{McQuinn2011}, which are also seen in four additional CMDs in Figure~\ref{fig:cmds} (AGC~110482, AGC~111977, AGC~174585, AGC~731457). The nature of the star formation histories in these galaxies based on their stellar populations will be investigated in a companion paper (McQuinn et al. in prep).

\section{Distance Determinations\label{distances}}
The presence of the RGB in the photometry of the SHIELD galaxies allows us to use the TRGB method to measure the distances. Briefly, the TRGB distance method is a standard candle approach that arises from the stable and predictable I band luminosity of low-mass stars just prior to the helium flash \citep{Mould1986, Freedman1988, DaCosta1990}. Photometry in the I band is preferred because the metallicity dependence is reduced compared to that of other optical bands \citep[e.g.,][]{Lee1993}. As a discontinuity in the I band LF identifies the TRGB luminosity, the TRGB detection method requires observations of resolved stellar populations without significant crowding effects in both the F606W filter (i.e., an HST V band equivalent) and the F814W filter (i.e., the HST I band equivalent), that reach $>1$ mag below the TRGB in the F814W data. In addition to a zero-point calibration, a correction for the metallicity dependence of the TRGB luminosity can be applied. We use the TRGB luminosity calibration from \citet{Rizzi2007} for the ACS filters which includes a color correction to account for varying metallicity in the stellar populations: 

\begin{align}
& M_{F814W}^{ACS} = -4.06(\pm0.02) \nonumber \\
& + 0.20(\pm0.01) \cdot [(F606W-F814W) - 1.23] \label{eq:trgb}
\end{align}

\subsection{Detection of the TRGB}
We used two techniques to measure the luminosity of the TRGB in the F814W photometry. First, we applied a Sobel filter edge detection technique to identify the discontinuity in the LF. However, the results from this approach are affected by the binning of the luminosity function and have greater uncertainties when applied to photometry reaching $\sim1$ mag below the TRGB \citep{Makarov2006}. To reduce uncertainties, we used a second method employing a Maximum Likelihood (ML) technique, which takes into account the photometric uncertainties and completeness of the photometry. 

Both techniques require that point sources be pre-selected from the photometry based on color and magnitude ranges of the RGB stars. However, the photometry in the F606W filter is not as deep as the F814W filter. Therefore, if uniform photometric cuts are applied to both filters, completeness effects are introduced from the bluer filter for sources with higher, and still acceptable, SNRs in the redder filter. Thus, as mentioned in \S2, we used more liberal SNR cuts in the F606W photometry lists for identifying which point sources to be used in the TRGB measurements. Once identified, only the F814W LF is used to measure the TRGB luminosity. As noted above, Figure~\ref{fig:cmd_Vsnr2_5} shows the photometry for AGC~174605 created from the F814W photometry with a SNR$>5\sigma$ and the F606W photometry with a SNR$>2.5\sigma$. The dashed black line encompasses the point sources used for the determination of the TRGB luminosity. Similar regions outlining the magnitude and color ranges for the rest of the sample are shown in the CMDs in Figure~\ref{fig:cmds}. 

\subsubsection{Sobel Filter}
The first method determines the TRGB luminosities by identifying the break in the F814W LF using a Sobel edge detection filter \citep{Lee1993, Sakai1996, Sakai1997}. The bin width of the F814W LF was varied from 0.05 mag to 0.10 mag depending on the quality of the data. The choice of bin size affects the measured uncertainties in the distances and is discussed in more detail below. A Sobel edge detection filter \citep{Sakai1996} was then applied to the F814W LF of the selected stars. Figure~\ref{fig:lf_sobel} shows the LFs (solid lines) and Sobel filter responses (dotted lines) from the data. The dashed red lines mark the luminosities identified as the TRGB discontinuities.

Note that the Sobel responses often have more than one significant peak. In order to determine which peak corresponds to the TRGB discontinuity, we checked each response against the data in the CMDs. For example, in AGC~112521, there are two peaks in the Sobel response which occur at F814W magnitudes of 24.95 and 25.38. The peak response at 24.95 mag corresponds to the TRGB identifiable by eye in the CMD, while the peak response at 25.38 mag lies below the TRGB feature in the CMD. 

The TRGB luminosities based on the Sobel filter responses and the uncertainties are shaded in red in the CMDs shown in Figure~\ref{fig:cmds}. The uncertainty in the Sobel filter responses are based on the half-width at half maximum  of the Sobel filter response. In galaxies like AGC~111164 that have well populated RGBs, photometry reaching nearly 2 mag below the TRGB, and few potentially non-RGB point sources near the TRGB in the CMD (i.e., from background contamination or asymptotic giant branch (AGB) stars), an F814W LF bin width of 0.05 mag yields a Sobel filter response with one significant peak that is fairly narrow. In this case, the uncertainty in the Sobel response is $\pm0.08$ mag. In other galaxies, like AGC~111946, that have photometry reaching only $\sim1$ mag below the TRGB and a larger number of potentially non-RGB point sources near the TRGB, an F814W bin width of 0.10 mag is needed to produce a Sobel filter response with a peak for the TRGB discontinuity. In this case, the uncertainty in the Sobel response is $\pm0.20$ mag. 

\subsubsection{Maximum Likelihood Technique}
The second method determines the TRGB luminosities by fitting a parametric RGB luminosity function to the observed distribution of stars \citep[e.g.,][]{Sandage1979, Mendez2002, Makarov2006}. The strength of this approach is that the probability estimation takes into account the photometric error distribution and completeness using artificial star tests \citep[see][for a full discussion]{Makarov2006}. For the theoretical LF, we assumed the following form:

\begin{subequations}
\begin{empheq}[left={P = }\empheqlbrace]{alignat=2}
	& 10^{(A*(m - m_{TRGB}) + B)}, & \quad \text{if m - m$_{TRGB} \geq 0$}\\
	& 10^{(C*(m - m_{TRGB}))}, & \quad \text{if m - m$_{TRGB} < 0$}
\end{empheq}
\label{eq:ml_form}
\end{subequations}

\noindent where A has a normal prior of 0.30 and $\sigma=0.07$; C has a normal prior of 0.30 and $\sigma=0.2$. This is the same theoretical LF form used in \citet{Makarov2006}. A and C were treated as free parameters for most of the sample, with the exceptions of AGC~731457 and AGC~749237. In these two cases, the photometry was not deep enough to constrain the RGB LF and, thus, the values of the LF slopes were fixed at 0.3. The returned value of $m_{TRGB}$ of the most likely solution is the best fit to the data. The range in solutions returning log(P) within 0.5 of the maximum gives the uncertainty (as is the case with a normal distribution). The results of the ML technique are overplotted as solid black lines on the CMDs in Figure~\ref{fig:cmds}, and solid red lines on the LFs in Figure~\ref{fig:lf_sobel}. 

\begin{figure}[h]
\includegraphics[width=\columnwidth]{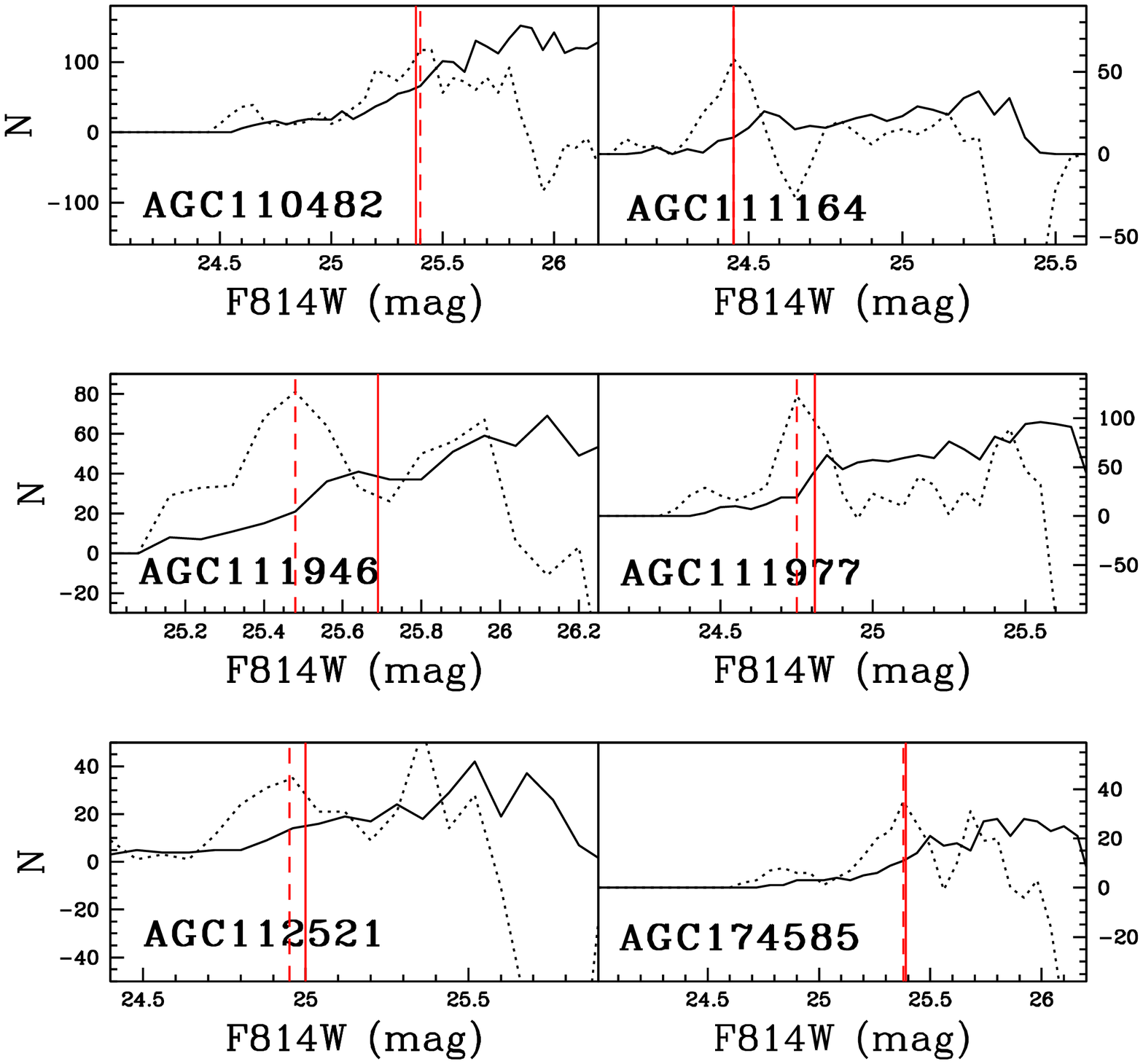}
\caption{F814W LF (solid line) and Sobel filter response (dotted line) of the galaxies. The dashed red line marks the break in the F814W LF identified by the Sobel edge detection algorithm. The solid red line marks the luminosity of the TRGB found by the maximum likelihood technique.}
\label{fig:lf_sobel}
\end{figure}

\begin{figure}
\addtocounter{figure}{-1}
\includegraphics[width=\columnwidth]{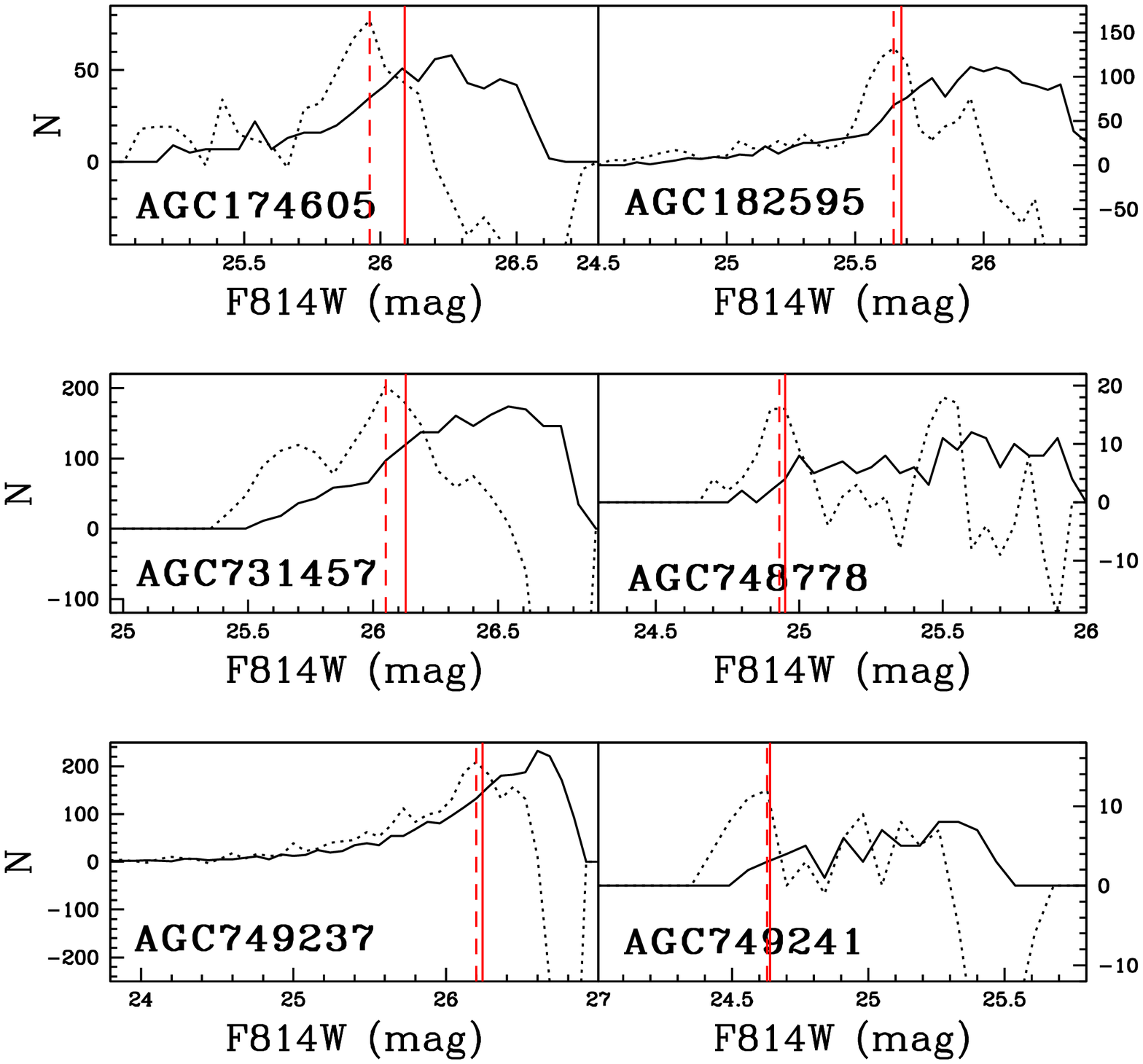}
\caption{\textit{Continued.}}
\end{figure}

\subsubsection{Comparison of the Sobel Filter and Maximum Likelihood Results}
Table~\ref{tab:distances} lists the TRGB luminosities identified both by the Sobel filter and by the ML technique; Figure~\ref{fig:cmds} shows the Sobel filter results with uncertainties shaded in red and the ML result drawn as a solid black line. 

Overall, the results from both the Sobel filter and ML methods agree within the uncertainties. In all but 2 cases, the Sobel filter response returns a brighter magnitude (closer distance) for the TRGB. It is well established that Sobel filer results are dependent on the LF bin size and bin placement, and therefore, are generally biased towards bright RGB stars. This difference is small for the galaxies with photometry reaching $\sim2$ mag below the TRGB with lower photometric uncertainties. For example, TRGB luminosities measured by the Sobel filter and ML technique are in very close agreement for AGC~111164, AGC~111977, AGC~112521, AGC~174585, AGC~748778, and AGC~749241. The difference in TRGB luminosities measured by the two methods is larger for some of the systems with photometry only reaching $\sim1$ mag below the TRGB (e.g., AGC~111946, AGC~174605, and AGC~731457). In these cases, the higher photometric uncertainties degrade the statistical sampling of the LF and can effect the Sobel filter response. The same effect is not as problematic for the ML technique as photometric incompleteness is taken into account by the artificial star tests. 

As the ML technique does not rely on binning and includes photometric uncertainties and completeness, we adopt the TRGB luminosities from the ML technique for all galaxies.

\subsection{Calculating the Distance Moduli}
The TRGB luminosities identified by the ML technique were converted to distance moduli using Equation~\ref{eq:trgb} and the average F606W - F814W color of the stars in the TRGB region of the CMDs. Table~\ref{tab:distances} lists the magnitude of the identified break in the F814W LF, the average F606W-F814W color of the stars in the TRGB regions, the calculated distance moduli, and the corresponding distance for each galaxy. The uncertainties are based on adding in quadrature the uncertainties from the ML technique, photometry, artificial star tests, and TRGB calibration. The distances to the SHIELD galaxies range from 5 $-$ 12 Mpc. 

Table~\ref{tab:distances} lists previously reported distances to the galaxies derived using various distance methods. First, TRGB distances were previously measured for two galaxies,  AGC~111164 and AGC~111977, by \citet{Tully2006} using ACS observations and by \citet{Karachentsev2003} using WFPC2 observations. Our measured TRGB distances are in agreement with the values from \citet{Tully2006}. The slight fractional differences in the TRGB distances of 4\% for AGC~111164 and 8\% for AGC~111977 are due to the use of the different zero-point calibration of the TRGB luminosity without a metallicity dependency correction of $M_I=-4.05$ mag in \citet{Tully2006}. Our measured TRGB distances are somewhat larger than those derived from the WFPC2 data in \citet{Karachentsev2003}. The fractional differences in TRGB distances of 8\% for AGC~11164 and 21\% for AGC~111977 are due to the higher uncertainties from the shallower WFPC2 data and from differences in the applied zero-point calibrations. 

Second, distances were previously estimated to three galaxies, AGC~110482, AGC~111946 \citep{Karachentsev2004}, and AGC~112521 \citep{Zitrin2008}, based on membership to the NGC~672 galaxy group. The average distance to this group (D $\sim7.2$ Mpc) was previously used as a distance estimate for these three systems. Our measured TRGB distances improve upon these estimates with fractional differences ranging from $-10$\% to $+20$\%. 

Finally, distances to seven systems were estimated in the ALFALFA survey using a parametric multi-attractor flow-model developed by \citet{Masters2005} with further discussion presented in \citet{Yahil1985}, \citet{Tonry2000}, and \citet{Martin2010}. Preliminary flow-model distances to these systems were reported in \citet{Cannon2011} and are listed in parentheses in Table~\ref{tab:distances}. The model parameters were updated and final flow-model distances were reported in the ALFALFA catalogue \citep{Haynes2011} with uncertainties of 2.3 Mpc, listed without parentheses in Column 9 of Table~\ref{tab:distances}. Note that the flow-model distance to one galaxy,  AGC~749241, has a higher uncertainty as this system is located in the complicated velocity field surrounding the Coma~I cluster. While the flow-model distance to this galaxy is a close match to the TRGB distance, it is likely only coincidence. The fractional differences between the TRGB distances reported here and the flow-model distances reported in \citet{Haynes2011} range from 0\% to 73\%; where discrepant, the flow-model estimates are all closer. The method used to estimate the distance to one galaxy, AGC~749237, was changed from a flow-model approach to one based on potential membership to the CVnI group. While the heliocentric radial velocity of 372 km~$^{-1}$ for AGC~749237 is consistent with the velocities of members in this group, the TRGB distance reported here places the galaxy farther than the CVnI group.

The fact that the flow model distances are smaller for this subsample also means that the \HI\ mass estimates used in selecting the sample were underestimated. Thus, the use of these distances introduced a bias in the sample. In fact, with the exception of AGC 112521, the distance estimates used to construct the SHIELD sample were {\it all} underestimates relative to our new TRGB distances.  This should come as no surprise, in hindsight, due to the way the sample was selected.  Since the goal of SHIELD was to study ALFALFA galaxies with the lowest HI masses, it is natural that the sample selection would preferentially include objects whose large distance uncertainties biased their distances low.   While this bias does mean that the properties of the SHIELD galaxies are not as extreme as originally thought, it does not negate the integrity of this study since the sample galaxies truly are low HI mass systems (see below).  This should serve as a caution against adopting flow-model distances, especially for nearby galaxies, since the differences in distance between the two methods are in several cases substantially larger than the 2.3 Mpc uncertainty indicated in \citet{Haynes2011}.

Table~\ref{tab:distances} also lists the \HI\ masses of the galaxies based on the ALFALFA fluxes reported in \citet{Cannon2011} and our reported distances. The \HI\ masses range from $4\times10^6$ to $6\times10^7$ \msun, with a median \HI\ mass of $1\times 10^7$ \msun\ and 5 galaxies with an \HI\ mass below $10^7$ \msun. This range of \HI\ masses overlaps with previous studies on low mass galaxies. For example, the FIGGS sample has an \HI\ mass range of $2\times10^6$ to $8\times10^8$ \msun, with a median \HI\ mass of $3\times 10^7$ \msun\ and 8 galaxies with an \HI\ mass below $10^7$ \msun\ \citep{Begum2008}. The LITTLE THINGS sample has an \HI\ mass range of $2\times10^5$ to $7\times10^8$ \msun, with a median \HI\ mass of $9\times 10^7$ \msun\ and 6 galaxies with an \HI\ mass below $10^7$ \msun\ \citep{Hunter2012}. Finally, the VLA-ANGST sample has an \HI\ mass range of $1\times10^5$ to $1\times10^9$ \msun, with a median \HI\ mass of $2\times 10^7$ \msun\ and 15 galaxies with an \HI\ mass below $10^7$ \msun; 6 of these are upper limits \citep{Ott2012}.

\section{Nearest Neighbors to the SHIELD Galaxies\label{neighbors}}

\begin{figure*}
\includegraphics[width=\textwidth]{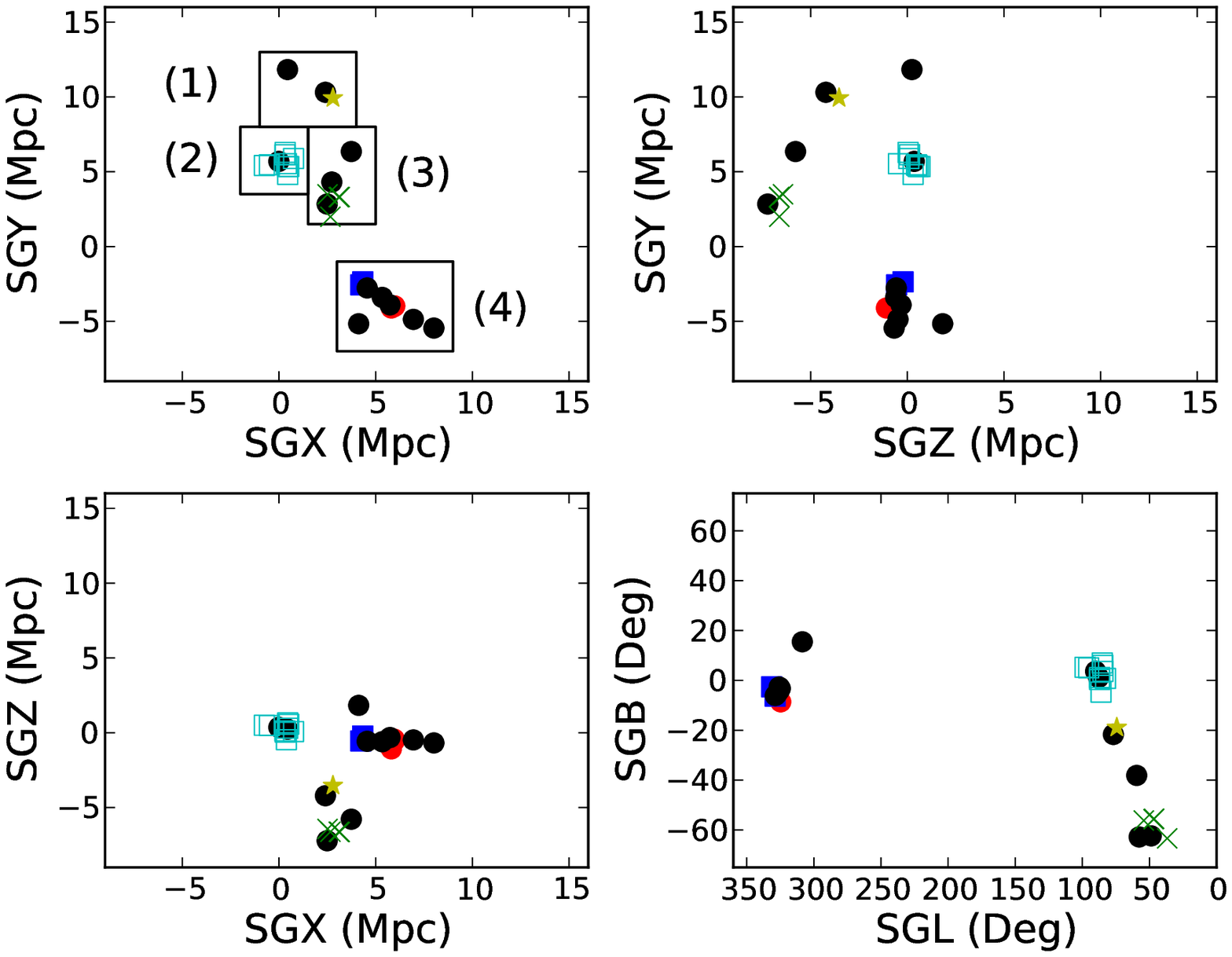}
\caption{Super Galactic (SG) coordinates show the 3-D distribution of the SHIELD sample relative to known galaxies, and help trace the distribution of matter in low luminosity environments with a low concentration of galaxies. Plotted in the first three panels are the SGX, SGY, and SGZ coordinates of the SHIELD galaxies (filled black circles), nearest neighbors to AGC~731457 (yellow star), AGC~749241 (unfilled cyan squares), AGC~174585 (green crosses), the NGC~784 group (filled blue squares), and the NGC~672 group (filled red circles). The final panel shows the same galaxies in SGL and SGB. The galaxies in the surrounding environments of the SHIELD sample were identified based on a spatial algorithm search using the Updated Nearby Galaxy Catalog \citep{Karachentsev2013}. The black boxes in the first panel highlight four regions that are shown in more detail in Figure~\ref{fig:subgroups}.}
\label{fig:all_groups}
\end{figure*}

As new distance measurements to low mass galaxies become available, the spatial distribution of galaxies populating the lower end of the galaxy luminosity function can provide important tracers of mass in low density regions. It has been estimated that approximately half of the galaxies in the LV reside in virialized groups and clusters \citep{Tully1987, Crook2007, Makarov2011}. Another 20\% are estimated to be located in the collapsing regions around groups and clusters \citep{Karachentsev2011}. The majority of dwarf galaxies not residing in and around groups or clusters are thought to exist in associations or bound structures of very low luminosity densities \citep{Tully2002, Tully2006}. In support of this result, a recent study of 10,900 galaxies with radial velocity measurements of $V_{LG} < 3500$ km s$^{-1}$, found that ``orphaned'' galaxies make up only 4.8\% of the galaxy population \citep{Karachentsev2011}. In other words, even though the LV is a relatively low density environment, few galaxies are thought to be truly isolated. 

Accurate distances to the SHIELD galaxies enable an identification of their nearest neighbors, furthering efforts to identify galaxy structures and trace the distribution of matter over a larger volume. Using the updated Nearby Galaxy Catalog \citep{Karachentsev2013}, we have searched for galaxies that are within a radius of 1 Mpc around the SHIELD sample in Supergalactic (SG) coordinates. Because the SG spherical coordinate system has its equator aligned with the major planar structure in the local universe \citep{deVaucouleurs1953}, SG coordinates help visualize the 3-D spatial distribution of nearby galaxies relative to the planar structure. Particularly useful are the SGX, SGY, and SGZ coordinates, defined in the standard way:

\begin{subequations}
\begin{align}
SGX = Distance \cdot cos(SGL) \cdot cos(SGB) \\
SGY = Distance \cdot sin(SGL) \cdot cos(SGB) \\
SGZ = Distance \cdot sin(SGB)
\end{align}
\label{eq:sg_coordinates}
\end{subequations}

Figure~\ref{fig:all_groups} shows the 3-D distribution in SG coordinates of the SHIELD sample and known galaxies within a 1 Mpc radius of each system. The first three panels in Figure~\ref{fig:all_groups} plot the SGX, SGY, and SGZ coordinates. The final panel plots the sky projected SGL and SGB coordinates of these galaxies. The SHIELD galaxies are plotted as filled black circles. The rest of the galaxies are color coded according to either known membership in a group or simply as nearest neighbors to a SHIELD system as follows: NGC~784 group (filled blue squares), NGC~672 group (filled red circles), nearest neighbors to AGC~749237 (yellow star), AGC~749241 (unfilled cyan squares), AGC~174585 (green crosses). The boxes highlight four different regions in SG coordinates that encompass the SHIELD galaxies. Each of these regions is shown in more detail in Figure~\ref{fig:subgroups}. 

Figure~\ref{fig:subgroups} expands the SG plots from the four regions highlighted in Figure~\ref{fig:all_groups}. The galaxies are color coded as in Figure~\ref{fig:all_groups}. Note that some of the regions have multiple SHIELD galaxies. Thus, the volumes encompassed by each plot are different. Eight SHIELD galaxies are located in regions with other galaxies. Four systems, AGC~174605, AGC~182595, AGC~748778, and AGC~749237 have no known galaxies within a radius of 1 Mpc. We summarize the environments around each of the SHIELD galaxies in detail below.

\subsection{The Environment Around AGC~731457}
\noindent AGC~731457 is located in region (1) in Figure~\ref{fig:all_groups} and the first plot of Figure~\ref{fig:subgroups}. This system has one neighbor, DDO~83, located 0.9 Mpc away. The heliocentric velocity of DDO~83 is 582 km~s$^{-1}$, compared with a velocity of 454 km~s$^{-1}$ for AGC~731457. Given the large physical separation and difference in velocities, it is unlikely that these galaxies are associated. At the greater distance of these systems ($\sim11$ Mpc), there may be additional low-mass systems that have yet to be identified or lack reliable distance measurements.

\begin{figure}
\includegraphics[width=\columnwidth]{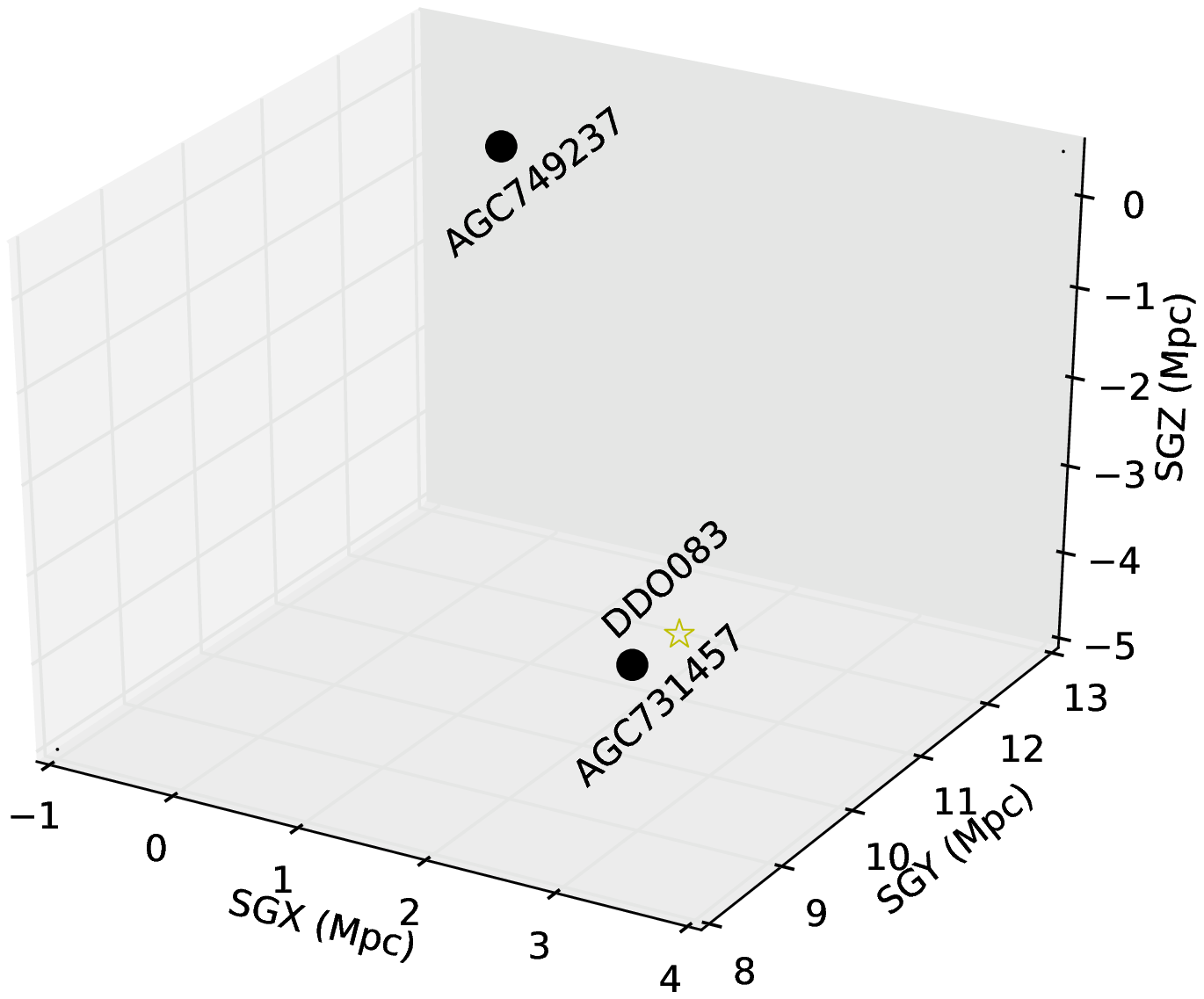}
\caption{Region (1) from Figure~\ref{fig:all_groups} is shown in more detail with the same colors and plot symbols. AGC~749237 had no known systems with 1 Mpc. AGC~731457 has one neighbor, DDO~83, located $\sim0.7$ Mpc away.}
\label{fig:subgroups}
\end{figure}

\subsection{The Environment Around AGC~749241}
\noindent AGC~749241 is located in region (2) in Figure~\ref{fig:all_groups} and the second plot of Figure~\ref{fig:subgroups}. This system resides in a more highly populated region with 8 low mass galaxies and one more massive spiral galaxy, NGC~4656. NGC~4656 has been identified as a member of the 14$-$6 group \citep{Tully1988, Tully2008}. The nearest neighbor to AGC~749241 is IC~3840 at a distance of 0.5 Mpc. IC~3840 has an heliocentric radial velocity of 557 km~s$^{-1}$, compared with a velocity of 451 km~s$^{-1}$ for AGC~749241. 

\begin{figure}
\figurenum{\ref{fig:subgroups}}
\includegraphics[width=\columnwidth]{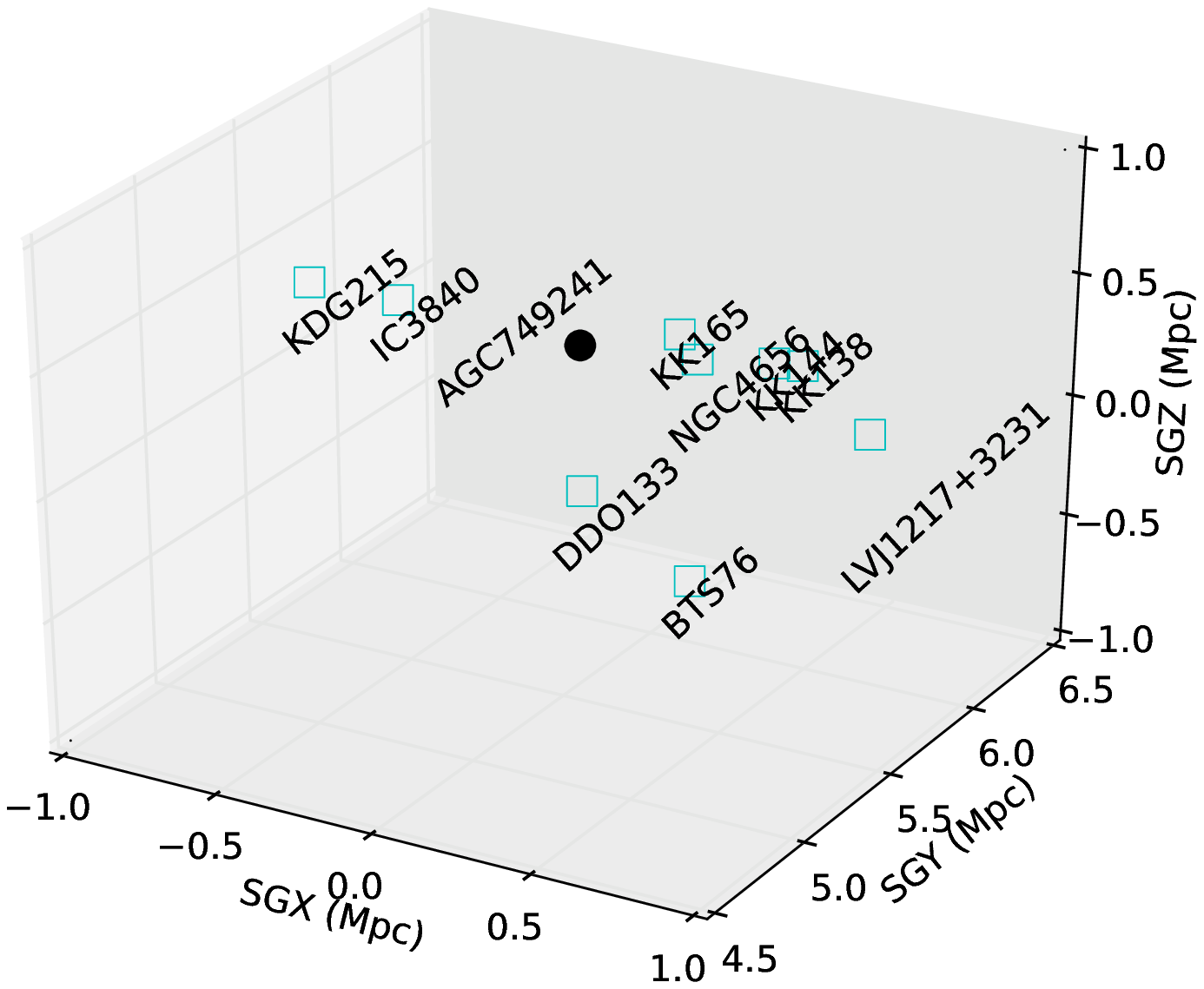}
\caption{\textit{Continued.} Region (2) from Figure~\ref{fig:all_groups} is shown in more detail with the same colors and plot symbols. AGC~749241 has 9 systems within 1 Mpc; its nearest neighbor is IC~3840 located $\sim0.5$ Mpc away. }
\end{figure}

The string of galaxies ranging from KDG~215 to LVJ1217$+$3231 lie along the same plane in SG coodinates creating a linear structure reaching 1.6 Mpc from end to end. The radial velocities of these galaxies range from 419 km~s$^{-1}$ for KDG~415 to 630 km~s$^{-1}$ for NGC~4656; four of the galaxies have velocities clustered between 419 km~s$^{-1}$ and 480 km~s$^{-1}$. A likely explanation of the tight linear structure is that these systems are arranged along a dark matter filament.

\subsection{The Environment Around AGC~174585}
\noindent AGC~174585 is located in region (3) in Figure~\ref{fig:all_groups} and the third plot of Figure~\ref{fig:subgroups}. This system does not have any close association with other systems, but lies $\sim0.9$ Mpc from the previously identified association, $14+19$, which includes DDO~47, KK~65, UGC~4115, and UGC~3755 \citep{Tully2006}. The \HI\ velocities of the members of the 14$+$19 range from 272$-$341 km~s$^{-1}$, compared with a velocity of 356 km~s$^{-1}$ for AGC~174585. Given the large physical separation, it is unlikely that AGC~174585 is part of the $14+19$ association.

\begin{figure}
\figurenum{\ref{fig:subgroups}}
\includegraphics[width=\columnwidth]{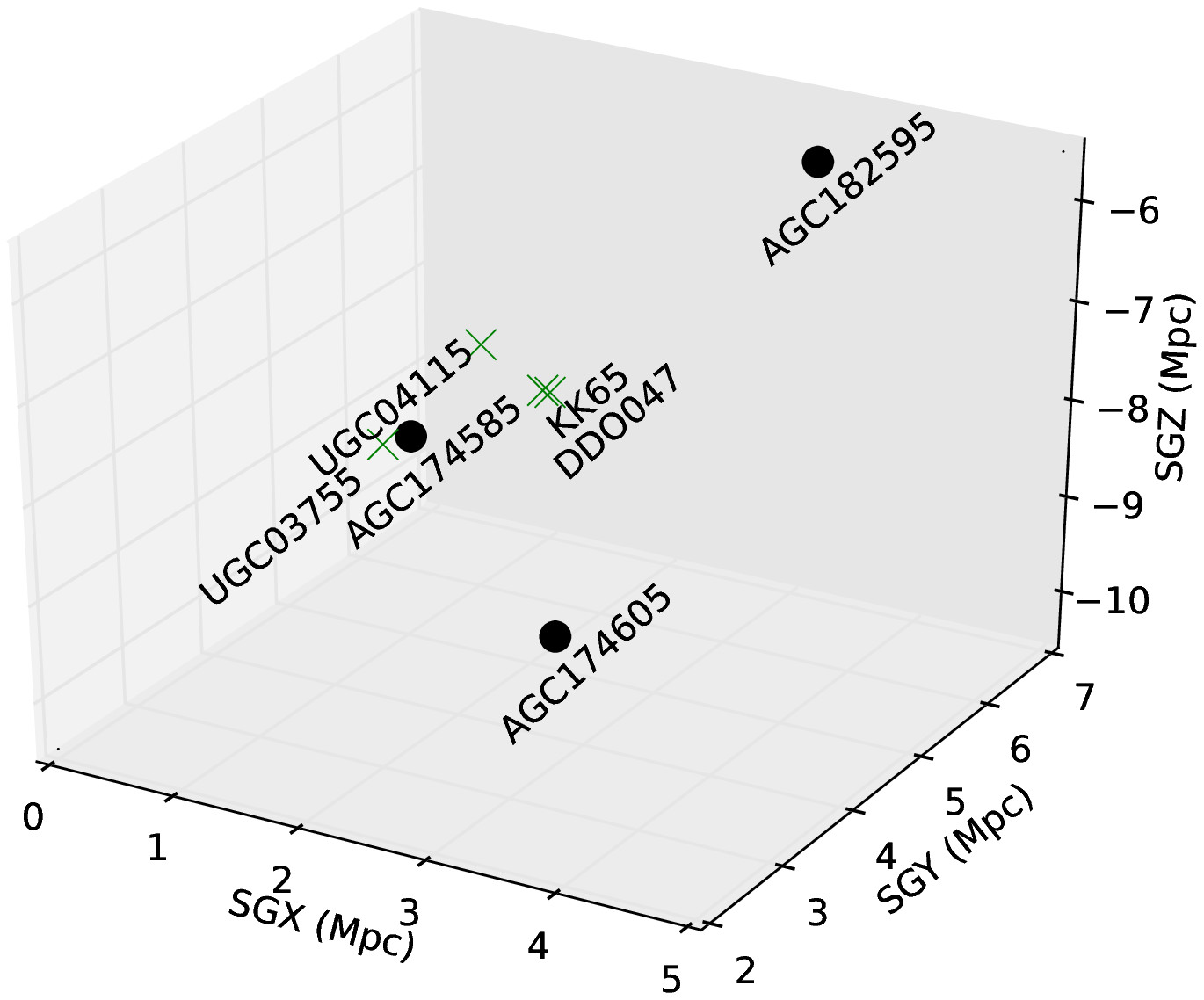}
\caption{\textit{Continued.} Region (3) from Figure~\ref{fig:all_groups} is shown in more detail with the same colors and plot symbols. AGC~174585 is located $\sim0.9$ Mpc away from the known association $14+19$. AGC~174605 and AGC~182595 have no galaxies within a radius of 1 Mpc; these are truly isolated galaxies.}
\end{figure}

\subsection{Truly Isolated: AGC~174605, AGC~182595, AGC~748778, AGC~749237}
\noindent AGC~749237 is located in region (1), AGC~174605 and AGC~182595 are located in region (3), and AGC~748778 is located in region (4) in Figure~\ref{fig:all_groups}. No known galaxies were identified within a 1 Mpc radius of these four galaxies. The four SHIELD galaxies have properties (i.e., late morphological type, low luminosity, significant gas content) consistent with other known isolated galaxies \citep{Karachentsev2011}. The closest systems identified are as follows:

\begin{itemize}
\item AGC~174605: Closest neighbors are the pair of galaxies DDO~47 and KK~65 at a distance of 3.2 Mpc that are part of the $14+19$ association.
\item AGC~182585: Closest neighbor is the galaxy LSBCD564$-$0 at a distance of 1.3 Mpc.
\item AGC~748778: Closest neighbor is the galaxy UGC~00685 at a distance of 2.4 Mpc.
\item AGC~749237: Closest neighbor is the galaxy IC~3308 at a distance of just over 1 Mpc.
\end{itemize}

Note that three of these systems are at distances greater than 10 Mpc. Thus, their isolation may be due to our incomplete knowledge of the galaxy population at these larger distances. 

\subsection{The Environment Around  AGC~110482, AGC~111164, AGC~111946, AGC~111977, AGC~112521: The NGC~672 and NGC~784 Groups}

\noindent There are six SHIELD galaxies located in region (4) in Figure~\ref{fig:all_groups} and the final plot of Figure~\ref{fig:subgroups}. With the exception of AGC~748778 (discussed in the previous subsection), all of the SHIELD systems have been previously associated with two galaxy groups, namely the NGC~672 and NGC~784 groups \citep[e.g.,][]{Karachentsev2004}. Specifically, AGC~110482, AGC~111946, and AGC~112521 have been identified as part of the NGC~672 group. The heliocentric radial velocities of the 3 SHIELD galaxies range from 274$-$367 km~s$^{-1}$ compared with the higher velocity of 429 km~s$^{-1}$ for NGC~672. AGC~111164 and AGC~111977 have been identified as part of the NGC~784 group. The radial velocities of these two SHIELD galaxies are 163 and 207  km~s$^{-1}$, which bracket the measured 193  km~s$^{-1}$ velocity of NGC~784. In addition, AGC~111945, another low-mass galaxy detected in the ALFALFA survey with a stellar component identified in SDSS imaging, has been previously linked to these 2 groups of galaxies. This galaxy has a very similar \HI\ mass and optical luminosity as the SHIELD sample \citep[cf.,][]{Cannon2011}. Finally, two other systems, AGC~122834, a small (i.e., 15\arcsec) dIrr galaxy, and AGC~122835, an \HI\ cloud with no identified optical counterpart, have been identified as possible members of the NGC~784 group \citep{Zitrin2008}. However, as these systems lack secure distances, we do not include them in Figures~\ref{fig:all_groups} or \ref{fig:subgroups}. 

\begin{figure}
\figurenum{\ref{fig:subgroups}}
\includegraphics[width=\columnwidth]{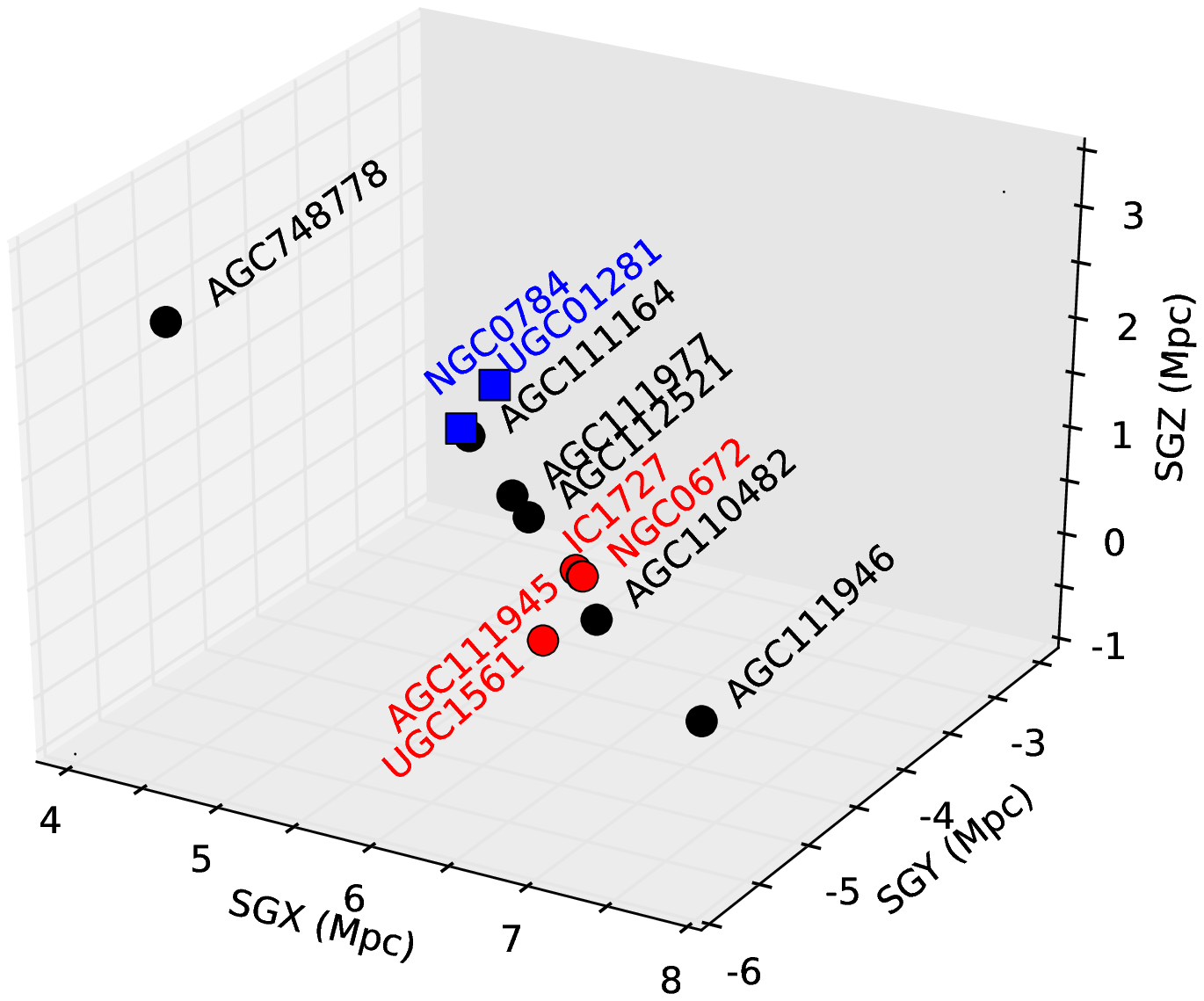}
\caption{\textit{Continued.} Region (4) from Figure~\ref{fig:all_groups} is shown in more detail with the same colors and plot symbols. AGC~111164, AGC~111977, AGC~112521, AGC~110482, and AGC~111946 appear to be part of two groups, NGC~784 and NGC~672. Together, these systems make up a single structure. Additional small systems have been previously associated with these groups, but lack secure distances and are thus not included in the Figure. The total linear structure spans just over 4~Mpc from UGC~01281 to AGC~111946. Also plotted, but not associated with this structure, is AGC~748778. AGC~748778 is an isolated galaxy with no neighbors identified within a 1 Mpc volume.}
\end{figure}

Our new distances revise the spatial distribution of the 5 SHIELD galaxies relative to the larger group members, and places AGC~111946 at a farther distance from NGC~672. Together, these galaxies create a clear structure in SG coordinates connecting the NGC 784 and NGC~672 groups.  A previous study of the two groups reported that the galaxies form a single kinematic ensemble \citep{Zitrin2008}. These authors suggest that the galaxies in both groups are located on a dark matter filament in a low-density galactic environment. The 3-D distribution of these 5 SHIELD galaxies and the other members of the NGC~672 and NGC~784 groups in Figure~\ref{fig:subgroups} supports the conclusion that these systems are part of a single kinematic structure, tracing the distribution of matter in this region. The total structure is highly linear, and extends just over 4~Mpc from UGC~01281 to AGC~111946.

Many of these galaxies have blue optical colors indicative of recent star formation, including the more massive systems in the groups (e.g., NGC~672, IC~1727, and NGC~784), and the lower mass SHIELD galaxies. The starburst in NGC~784 has been studied in detail by \citet{McQuinn2010, McQuinn2012}. These authors report the star formation in NGC~784 is distributed across the optical disk of the galaxy and has been on-going for $450\pm50$ Myr. The star formation activity in these groups has been attributed both to interactions between galaxies \citep[e.g., NGC~672 and IC~1727;][]{Combes1980, Sohn1996} and to accretion along the filament \citep{Zitrin2008}. However, additional \HI\ observations obtained on the Green Bank Telescope have failed to detect any lower mass \HI\ clouds down to a 5$\sigma$ mass detection threshold of $6\times10^6$ \msun, in contradiction to the accretion model \citep{Chynoweth2011}.

\section{Conclusions \label{conclusions}}
We have used optical imaging of resolved stellar populations obtained from the $HST$ to measure the TRGB distance to 12 low-mass, gas-rich galaxies in the SHIELD program. The distances to the galaxies range from $5 - 12$ Mpc. The CMDs show a population of young main sequence stars indicating these systems have experienced recent star formation. A subset of the sample also shows red and blue helium burning sequences suggesting the recent star formation has been on-going for $\gtsimeq$ 100 Myr. The star formation histories will be investigated in detail in a future paper (McQuinn et al. in prep). Five galaxies are part of two galaxy groups, NGC~784 and NGC~672, which are likely a single, bound kinematic structure. AGC~174585 is located $\sim1.0$ Mpc away from the previously identified association of galaxies, $14+19$, but is likely unbound to this structure. Similarly, AGC~731457 lies 0.9 Mpc from another system, DDO~83. AGC~749241 is situated in a more populated volume, forming part of 1.6 Mpc linear structure with six other galaxies. Similarly to the NGC~672 and NGC~784 groups, these galaxies are likely bound, lying along a dark matter filament. The nearest neighbor identified is $\sim0.5$ Mpc away. Four galaxies, AGC~174605, AGC~182595, AGC~748778, and AGC~749237 appear to be truly isolated with no known neighbors identified within a radius of 1 Mpc, although three of these galaxies are at relatively large distances (i.e., $\gtsimeq10$ Mpc) where our knowledge of the galaxy population is incomplete.  

Based on the distances and the \HI\ measurements from the ALFALFA survey, the SHIELD galaxies have \HI\ masses ranging from $4\times10^6$ to $6\times10^7$ \msun, with a median \HI\ mass of $1\times 10^7$ \msun\ and 5 galaxies with an \HI\ mass below $10^7$ \msun. Thus, the SHIELD program probes the low mass regime of gas-rich galaxies in a larger cosmological volume than previous surveys. As these 12 galaxies were selected from the first $\sim10$\% of the reduced ALFALFA database, it suggests that a large sample of low luminosity galaxies in this mass range exists in the nearby universe. Nominally, one might expect the total yield from the ALFALFA survey to be 10$\times$ greater. However, investigation of the first 40\% of the ALFALFA catalogue \citep{Haynes2011} suggests the true yield of low luminosity galaxies with low \HI\ velocities and inferred masses is likely more than 12$\times$ greater.

\section{Acknowledgments}
Support for this work was provided by NASA through grant GO-12658 from the Space Telescope Institute, which is operated by Aura, Inc., under NASA contract NAS5-26555. JMC is supported by NSF grant AST-1211683. E.~D.~S. is grateful for partial support from the University of Minnesota. Partial support for publication charges was provided by the NRAO. The National Radio Astronomy Observatory is a facility of the National Science Foundation operated under cooperative agreement by Associated Universities, Inc. The authors acknowledge the work of the entire ALFALFA collaboration team in observing, flagging, and extracting the catalogue of galaxies used to identify the SHIELD sample. The ALFALFA team at Cornell is supported by NSF grant AST-0607007 to R.G. and M.P.H. and by a grant to M.P.H. from the Brinson Foundation.  This research made use of NASA's Astrophysical Data System and the NASA/IPAC Extragalactic Database (NED) which is operated by the Jet Propulsion Laboratory, California Institute of Technology, under contract with the National Aeronautics and Space Administration. The authors acknowledge helpful discussions with Brad Jacobs and insightful comments from Noah Brosch. Finally, the authors thank the anonymous referee for constructive comments which helped improve this work. 

{\it Facilities:} \facility{Hubble Space Telescope}


\begin{deluxetable}{lccccrrrrrrcc}
\tablewidth{0pt}
\tablecaption{Summary of SHIELD Sample and Observations\label{tab:galaxies}}
\tablecolumns{13}
\tablehead{
\colhead{}     			&
\colhead{}			&
\colhead{R.A.}         		&
\colhead{Decl.}		     	&
\colhead{V$_{21}$}		&
\colhead{HST}			&
\colhead{F606W}       		&
\colhead{F814W}         	&
\colhead{a}			&
\colhead{1-$^b/_a$}		&
\colhead{P.A.}			&
\colhead{$A_{F606W}$}		&
\colhead{$A_{F814W}$}			
\\
\colhead{Galaxy}       		&
\colhead{Alt. Name}		&
\colhead{J2000}         	&
\colhead{J2000}         	&
\colhead{km s$^{-1}$}		&
\colhead{ID} 			&
\colhead{(sec)}       		&
\colhead{(sec)}	 		&
\colhead{($\arcsec$)}		&
\colhead{}			&		
\colhead{(\degree)}		&
\colhead{(mag)}			&
\colhead{(mag)}
\\
\colhead{(1)}      		&       
\colhead{(2)}        		&       
\colhead{(3)}      		&       
\colhead{(4)}        		&       
\colhead{(5)}        		&       
\colhead{(6)}          		&        
\colhead{(7)}          		&   
\colhead{(8)}          		&   
\colhead{(9)}			&                              
\colhead{(10)}                  &           
\colhead{(11)}			&                              
\colhead{(12)}			&                              
\colhead{(13)}                              
}
\startdata
AGC~110482  & KK~13 & 01:42:17.4 & 26:22:00 & 357 & 12658 & 1008 & 1226 & 37 & 0.30 & 15 & 0.26 & 0.18\\
AGC~111164  & KK~17 & 02:00:10.1 & 28:49:52 & 163 & 10210 &  934 & 1226 & 60 & 0.40 &  6 & 0.16 & 0.11\\      
AGC~111946  & KK~15 & 01:46:42.2 & 26:48:05 & 367 & 12658 & 1008 & 1126 & 27 & 0.56 &  6 & 0.24 & 0.16\\     
AGC~111977  & KK~16 & 01:55:20.2 & 27:57:14 & 207 & 12658 & 1008 & 1226 & 70 & 0.47 & 30 & 0.20 & 0.14\\     
	    &       & 		 & 	    &     & 10210 &  934 & 1226 &    &      &    &	  &	\\
AGC~112521  &       & 01:41:07.6 & 27:19:24 & 274 & 12658 & 1008 & 1226 & 32 & 0.38 & 18 & 0.18 & 0.12\\     
AGC~174585  &       & 07:36:10.3 & 09:59:11 & 356 & 12658 & 1000 & 1220 & 23 & 0.26 &347 & 0.11 & 0.07\\    
	    &       & 		 & 	    &     & 10210 &  900 &  900 &    &      &    &	  &	\\
AGC~174605  &       & 07:50:21.7 & 07:47:40 & 351 & 12658 & 1000 & 1200 & 26 & 0.19 &  0 & 0.07 & 0.05\\ 
AGC~182595  &       & 08:51:12.1 & 27:52:48 & 398 & 12658 & 1008 & 1226 & 52 & 0.19 &  0 & 0.12 & 0.08\\     
AGC~731457  &       & 10:31:55.8 & 28:01:33 & 454 & 12658 & 1008 & 1226 & 37 & 0.38 &  0 & 0.08 & 0.05\\      
AGC~748778  &       & 00:06:34.3 & 15:30:39 & 258 & 12658 & 1008 & 1226 & 27 & 0.11 &  0 & 0.19 & 0.13\\      
AGC~749237  &       & 12:26:23.4 & 27:44:44 & 372 & 12658 & 1008 & 1226 & 38 & 0.45 & 44 & 0.06 & 0.04\\      
AGC~749241  &       & 12:40:01.7 & 26:19:19 & 451 & 12658 & 1008 & 1226 & 27 & 0.07 &  0 & 0.04 & 0.03\\
\enddata
\tablecomments{\scriptsize{Column 1$-$Galaxy name. Column 2$-$Alternate galaxy name. Columns 3 and 4$-$Coordinates of galaxy in J2000. Column 5$-$ Heliocentric velocity from \citet{Haynes2011}. Column 6$-$HST observing program. Columns 7 and 8$-$Integration time in the F606W and F814W filters with the ACS instrument. Column 9$-$Semi-major axis. Column 10$-$Ellipticity. Column 11$-$Position angles. Elliptical parameters were empirically determined based on the CMDs of the stellar populations. See text for details. Columns 12 and 13$-$Galactic absorption from the dust maps of \citet{Schlegel1998}}}

\end{deluxetable}


\begin{turnpage}
\begin{deluxetable}{lccccrccrcrr}
\tabletypesize{\tiny}
\tablewidth{0pt}
\tablecaption{Distances to SHIELD Galaxies\label{tab:distances}}
\tablecolumns{12}
\tablehead{
\colhead{}			&
\colhead{Sobel Filter}		&
\colhead{ML}			&
\colhead{F606W-F814W}		&
\colhead{Distance}		&
\colhead{}			&
\colhead{}			&
\colhead{log}			&
\colhead{Prev.}			&
\colhead{}			&
\colhead{}			&
\colhead{}			
\\
\colhead{}     			&
\colhead{F814W$_{TRGB}$}   	&
\colhead{F814W$_{TRGB}$}   	&
\colhead{of TRGB}	    	&
\colhead{Modulus}		&
\colhead{Distance}       	&
\colhead{S$_{HI}$}		&
\colhead{\HI\ Mass}		&
\colhead{Dist.}			&
\colhead{}			&
\colhead{}			&
\colhead{Fractional}			
\\
\colhead{Galaxy}       		&
\colhead{(mag)}         	&
\colhead{(mag)}         	&
\colhead{(mag)}         	&
\colhead{(mag)} 		&
\colhead{(Mpc)}       		&
\colhead{(Jy~km~s$^{-1}$)}	&
\colhead{(\msun)}		&
\colhead{(Mpc)}			&
\colhead{Method}		&
\colhead{Ref.}			&
\colhead{Diff.}					
\\
\colhead{(1)}      		&       
\colhead{(2)}        		&       
\colhead{(3)}      		&       
\colhead{(4)}        		&       
\colhead{(5)}        		&
\colhead{(6)}        		&       
\colhead{(7)}        		&
\colhead{(8)}			&      
\colhead{(9)}        		&
\colhead{(10)}			&
\colhead{(11)}			&
\colhead{(12)}
}
\startdata
AGC~110482  & 25.40$\pm0.16$ & 25.38$\pm0.05$		& 1.10 & 29.47$\pm0.05$ 		& 7.82$\pm0.21$ 	& 1.33 & 7.28 & 7.2  	   & mem  & 1 &  8\%  \\
AGC~111164  & 24.45$\pm0.08$ & 24.45$\pm0.02$		& 1.08 & 28.54$\pm0.03$ 		& 5.11$\pm0.07$ 	& 0.65 & 6.61 & 4.9$\pm0.2$& TRGB & 2 &  4\%  \\
	    &		     &	    			&      &     				&	      		&      &      & 4.7$\pm0.5$& TRGB & 3 &  8\%  \\
AGC~111946  & 25.48$\pm0.20$ & 25.69$_{-0.06}^{+0.04}$	& 1.10 & 29.78$_{-0.06}^{+0.05}$ 	& 9.02$_{-0.29}^{+0.20}$& 0.76 & 7.17 & 7.2  	   & mem  & 1 & 20\%  \\
AGC~111977  & 24.75$\pm0.10$ & 24.81$_{-0.02}^{+0.03}$	& 1.20 & 28.88$_{-0.03}^{+0.04}$ 	& 5.96$_{-0.09}^{+0.11}$& 0.85 & 6.85 & 5.5$\pm0.1$& TRGB & 2 &  8\%  \\
	    &		     &	    			&      &	        		&	      		&      &      & 4.7$\pm0.5$& TRGB & 3 & 21\%  \\
AGC~112521  & 24.95$\pm0.20$ & 25.00$\pm0.05$		& 1.08 & 29.09$\pm0.05$ 		& 6.58$\pm0.18$ 	& 0.69 & 7.11 & 7.2 	   & mem  & 4 &$-$10\%\\
AGC~174585  & 25.38$\pm0.16$ & 25.39$_{-0.04}^{+0.05}$	& 1.05 & 29.49$\pm0.05$ 		& 7.89$_{-0.17}^{+0.21}$& 0.54 & 6.90 & (6.1) 5.0  & flow & (5) 6 & (23\%) 37\% \\
AGC~174605  & 25.96$\pm0.20$ & 26.09$\pm0.05$		& 1.05 & 30.19$\pm0.05$ 		&10.89$\pm0.28$ 	& 0.66 & 7.27 & (6.0) 4.8  & flow & (5) 6 & (45\%) 56\%\\
AGC~182595  & 25.65$\pm0.12$ & 25.68$\pm0.06$		& 1.05 & 29.78$\pm0.06$ 		& 9.02$\pm0.28$ 	& 0.42 & 7.00 & (6.1) 5.9  & flow & (5) 6 & (32\%) 35\%\\
AGC~731457  & 26.05$\pm0.17$ & 26.13$_{-0.02}^{+0.03}$	& 1.02 & 30.23$_{-0.03}^{+0.04}$	&11.13$_{-0.16}^{+0.20}$& 0.62 & 7.26 & (5.4) 6.1  & flow & (5) 6 & (52\%) 45\%\\
AGC~748778  & 24.93$\pm0.11$ & 24.95$_{-0.05}^{+0.04}$ 	& 1.03 & 29.05$\pm0.05$ 		& 6.46$_{-0.17}^{+0.14}$& 0.46 & 6.67 & (5.4) 4.6  & flow & (5) 6 & (16\%) 29\%\\
AGC~749237  & 26.20$\pm0.17$ & 26.24$_{-0.02}^{+0.03}$	& 1.10 & 30.33$_{-0.03}^{+0.04}$ 	&11.62$_{-0.16}^{+0.20}$& 1.80 & 7.76 & 3.2 	   & flow &  5    &  73\% \\
	    &		     &	    			&      &	      			&      	      		&      &      & 7.0	   & mem  &  6    &  40\% \\
AGC~749241  & 24.63$\pm0.13$ & 24.64$_{-0.05}^{+0.06}$	& 0.98 & 28.75$_{-0.05}^{+0.06}$ 	& 5.62$_{-0.14}^{+0.17}$& 0.76 & 6.75 & (4.3) 5.6  & flow & (5) 6 & (24\%)  0\%\\
\enddata
\tablecomments{Column 1$-$Galaxy name. Column 2$-$Identified break in the F814W LF corresponding to the magnitude of the TRGB using a Sobel Filter. The uncertainties were determined by the HWHM of the Sobel filter response. Column 3$-$The magnitude of the TRGB luminosity in the F814W LF from the ML approach. Column 4$-$Average F606W-F814W color of stars in the region of the TRGB. Column 5$-$Distance modulus calculated using Equation~\ref{eq:trgb}, the TRGB luminosity from the ML technique listed in Col. 3, and the average TRGB color from Col. 4. Uncertainties were determined by adding in quadrature the uncertainties of the calibration of the TRGB luminosity, and the uncertainties in the fit from the ML technique which take into account uncertainties from the photometry and artificial star tests. Column 6$-$ Distance to galaxy based on the distance modulus. Column 7$-$ALFALFA \HI\ flux integrals; uncertainties on the fluxes are of order 10\% \citep{Cannon2011}. Column 8$-$\HI\ mass based on the ALFALFA \HI\ flux measurements and the measured TRGB distances reported here. Column 9$-$Previous distance measurements to the galaxies. Preliminary flow-model distances from the ALFALFA survey were reported in \citet{Cannon2011} and are listed with parentheses; the final flow-model distances were reported in \citet{Haynes2011} and are listed without parenthesis. Column 10$-$Method used to determine the previous distance measurements based on membership to a known galaxy group (mem), the TRGB method (TRGB), or derivations from the parametric multi-attractor flow model (flow). Column 11$-$References for previous distances (listed below). Column 12$-$Fractional difference between the TRGB measured distances reported here in Column 5 and the previously measured distances listed in Column 9.}
\tablerefs{ 1-\citet{Karachentsev2004}; 2-\citet{Tully2006}; 3-\citet{Karachentsev2003}; 4-\citet{Zitrin2008}; 5-\citet{Cannon2011}; 6-\citet{Haynes2011}}

\end{deluxetable}
\end{turnpage}


\begin{thebibliography}{}
\bibitem[Anderson \& Bedin(2010)]{Anderson2010} Anderson, J., \& Bedin, L.~R.\ 2010, \pasp, 122, 1035 
\bibitem[Barnes et al.(2001)]{Barnes2001} Barnes, D.~G., 
Staveley-Smith, L., de Blok, W.~J.~G., et al.\ 2001, \mnras, 322, 486 
\bibitem[Begum et al.(2008)]{Begum2008} Begum, A., Chengalur, 
J.~N., Karachentsev, I.~D., Sharina, M.~E., 
\& Kaisin, S.~S.\ 2008, \mnras, 386, 1667 
\bibitem[Cannon et al.(2011)]{Cannon2011} Cannon, J.~M., 
Giovanelli, R., Haynes, M.~P., et al.\ 2011, \apjl, 739, L22 
\bibitem[Chynoweth et al.(2011)]{Chynoweth2011} Chynoweth, K.~M., 
Holley-Bockelmann, K., Polisensky, E., 
\& Langston, G.~I.\ 2011, \aj, 142, 137 
\bibitem[Combes et al.(1980)]{Combes1980} Combes, F., Foy, F.~C., Weliachew, L., \& Gottesman, S.~T.\ 1980, \aap, 84, 85 
\bibitem[Courtois et al.(2011a)]{Courtois2011a} Courtois, H.~M., 
Tully, R.~B., \& H{\'e}raudeau, P.\ 2011, \mnras, 415, 1935 
\bibitem[Courtois et al.(2011b)]{Courtois2011b} Courtois, H.~M., 
Tully, R.~B., Makarov, D.~I., et al.\ 2011, \mnras, 414, 2005 
\bibitem[Crook et al.(2007)]{Crook2007} Crook, A.~C., Huchra, 
J.~P., Martimbeau, N., et al.\ 2007, \apj, 655, 790 
\bibitem[Da Costa \& Armandroff(1990)]{DaCosta1990} Da Costa, G.~S., \& Armandroff, T.~E.\ 1990, \aj, 100, 162 
\bibitem[de Vaucouleurs(1953)]{deVaucouleurs1953} de Vaucouleurs, G.\ 
1953, \aj, 58, 30 
\bibitem[de Vaucouleurs(1975)]{deVaucouleurs1975} de Vaucouleurs, G.\ 
1975, Galaxies and the Universe, Editors Sandage, A., Sandage, M., and Kristian, J., University of Chicago Press, 557 
\bibitem[Dolphin(2000)]{Dolphin2000} Dolphin, A.~E.\ 2000, \pasp, 
112, 1383 
\bibitem[Dolphin(2009)]{Dolphin2009} Dolphin, A.~E.\ 2009, \pasp, 
121, 655 
\bibitem[Ford et al.(1998)]{Ford1998} Ford, H.~C., Bartko, F., 
Bely, P.~Y., et al.\ 1998, \procspie, 3356, 234 
\bibitem[Freedman(1988)]{Freedman1988} Freedman, W.~L.\ 1988, \apj, 
326, 691 
\bibitem[Giovanelli et al.(2013)]{Giovanelli2013} Giovanelli, R., 
Haynes, M.~P., Adams, E.~A.~K., et al.\ 2013, \aj, 146, 15 
\bibitem[Giovanelli et al.(2005)]{Giovanelli2005} Giovanelli, R., 
Haynes, M.~P., Kent, B.~R., et al.\ 2005, \aj, 130, 2598 
\bibitem[Girardi et al.(2010)]{Girardi2010} Girardi, L., Williams, 
B.~F., Gilbert, K.~M., et al.\ 2010, \apj, 724, 1030 
\bibitem[Haynes et al.(2011)]{Haynes2011} Haynes, M.~P., 
Giovanelli, R., Martin, A.~M., et al.\ 2011, \aj, 142, 170 
\bibitem[Hunter et al.(2012)]{Hunter2012} Hunter, D.~A., 
Ficut-Vicas, D., Ashley, T., et al.\ 2012, \aj, 144, 134 
\bibitem[Hunter et al.(1982)]{Hunter1982} Hunter, D.~A., 
Gallagher, J.~S., \& Rautenkranz, D.\ 1982, \apjs, 49, 53 
\bibitem[Karachentsev et al.(2004)]{Karachentsev2004} Karachentsev, 
I.~D., Karachentseva, V.~E., Huchtmeier, W.~K., 
\& Makarov, D.~I.\ 2004, \aj, 127, 2031 
\bibitem[Karachentsev et al.(2009)]{Karachentsev2009} Karachentsev, 
I.~D., Kashibadze, O.~G., Makarov, D.~I., 
\& Tully, R.~B.\ 2009, \mnras, 393, 1265 
\bibitem[Karachentsev et al.(2013)]{Karachentsev2013} Karachentsev, 
I.~D., Makarov, D.~I., \& Kaisina, E.~I.\ 2013, \aj, 145, 101 
\bibitem[Karachentsev et al.(2011)]{Karachentsev2011} Karachentsev, 
I.~D., Makarov, D.~I., Karachentseva, V.~E., 
\& Melnyk, O.~V.\ 2011, Astrophysical Bulletin, 66, 1 
\bibitem[Karachentsev et al.(2003)]{Karachentsev2003} Karachentsev, I.~D., Makarov, D.~I., Sharina, M.~E., et al.\ 2003, \aap, 398, 479 
\bibitem[Lee et al.(1990)]{Lee1990} Lee, Y.-W., Demarque, P., 
\& Zinn, R.\ 1990, \apj, 350, 155 
\bibitem[Lee et al.(1993)]{Lee1993} Lee, M.~G., Freedman, 
W.~L., \& Madore, B.~F.\ 1993, \apj, 417, 553 
\bibitem[Makarov \& Karachentsev(2011)]{Makarov2011} Makarov, D., \& Karachentsev, I.\ 2011, \mnras, 412, 2498 
\bibitem[Makarov et al.(2006)]{Makarov2006} Makarov, D., Makarova, 
L., Rizzi, L., et al.\ 2006, \aj, 132, 2729 
\bibitem[Martin et al.(2010)]{Martin2010} Martin, A.~M., 
Papastergis, E., Giovanelli, R., et al.\ 2010, \apj, 723, 1359 
\bibitem[Massey et al.(2010)]{Massey2010} Massey, R., Stoughton, 
C., Leauthaud, A., et al.\ 2010, \mnras, 401, 371 
\bibitem[Masters(2005)]{Masters2005} Masters, K.~L.\ 2005, 
Ph.D.~Thesis  
\bibitem[McQuinn et al.(2013)]{McQuinn2013} McQuinn, K.~B.~W., 
Skillman, E.~D., Berg, D., et al.\ 2013, \aj, 146, 145 
\bibitem[McQuinn et al.(2010)]{McQuinn2010} McQuinn, K.~B.~W., 
Skillman, E.~D., Cannon, J.~M., et al.\ 2010, \apj, 724, 49 
\bibitem[McQuinn et al.(2011)]{McQuinn2011} McQuinn, K.~B.~W., 
Skillman, E.~D., Dalcanton, J.~J., et al.\ 2011, \apj, 740, 48 
\bibitem[McQuinn et al.(2012)]{McQuinn2012} McQuinn, K.~B.~W., 
Skillman, E.~D., Dalcanton, J.~J., et al.\ 2012, \apj, 759, 77 
\bibitem[M{\'e}ndez et al.(2002)]{Mendez2002} M{\'e}ndez, B., 
Davis, M., Moustakas, J., et al.\ 2002, \aj, 124, 213 
\bibitem[Meyer et al.(2004)]{Meyer2004} Meyer, M.~J., Zwaan, 
M.~A., Webster, R.~L., et al.\ 2004, \mnras, 350, 1195 
\bibitem[Mould \& Kristian(1986)]{Mould1986} Mould, J., \& Kristian, J.\ 1986, \apj, 305, 591 
\bibitem[Ott et al.(2012)]{Ott2012} Ott, J., Stilp, A.~M., 
Warren, S.~R., et al.\ 2012, \aj, 144, 123 
\bibitem[Rhode et al.(2013)]{Rhode2013} Rhode, K.~L., Salzer, 
J.~J., Haurberg, N.~C., et al.\ 2013, \aj, 145, 149 
\bibitem[Rizzi et al.(2007)]{Rizzi2007} Rizzi, L., Tully, R.~B., 
Makarov, D., et al.\ 2007, \apj, 661, 815 
\bibitem[Sakai et al.(1996)]{Sakai1996} Sakai, S., Madore, B.~F., 
\& Freedman, W.~L.\ 1996, \apj, 461, 713 
\bibitem[Sakai et al.(1997)]{Sakai1997} Sakai, S., Madore, B.~F., 
\& Freedman, W.~L.\ 1997, \apj, 480, 589 
\bibitem[Sandage et al.(1979)]{Sandage1979} Sandage, A., Tammann, 
G.~A., \& Yahil, A.\ 1979, \apj, 232, 352 
\bibitem[Schlegel et al.(1998)]{Schlegel1998} Schlegel, D.~J., 
Finkbeiner, D.~P., \& Davis, M.\ 1998, \apj, 500, 525 
\bibitem[Skillman et al.(2013)]{Skillman2013} Skillman, E.~D., 
Salzer, J.~J., Berg, D.~A., et al.\ 2013, \aj, 146, 3 
\bibitem[Sohn \& Davidge(1996)]{Sohn1996} Sohn, Y.-J., \& Davidge, T.~J.\ 1996, \aj, 112, 2559 
\bibitem[Tonry et al.(2000)]{Tonry2000} Tonry, J.~L., Blakeslee, 
J.~P., Ajhar, E.~A., \& Dressler, A.\ 2000, \apj, 530, 625 
\bibitem[Tully(1987)]{Tully1987} Tully, R.~B.\ 1987, \apj, 321, 
280 
\bibitem[Tully(1988)]{Tully1988} Tully, R.~B.\ 1988, Cambridge 
and New York, Cambridge University Press, 1988, Nearby Galaxy Catalog, 221 
\bibitem[Tully \& Courtois(2012)]{Tully2012} Tully, R.~B., \& Courtois, H.~M.\ 2012, \apj, 749, 78 
\bibitem[Tully et al.(2013)]{Tully2013} Tully, R.~B., Courtois, 
H.~M., Dolphin, A.~E., et al.\ 2013, arXiv:1307.7213 
\bibitem[Tully et al.(2006)]{Tully2006} Tully, R.~B., Rizzi, L., 
Dolphin, A.~E., et al.\ 2006, \aj, 132, 729 
\bibitem[Tully et al.(2009)]{Tully2009} Tully, R.~B., Rizzi, L., 
Shaya, E.~J., et al.\ 2009, \aj, 138, 323 
\bibitem[Tully et al.(2008)]{Tully2008} Tully, R.~B., Shaya, 
E.~J., Karachentsev, I.~D., et al.\ 2008, \apj, 676, 184 
\bibitem[Tully et al.(2002)]{Tully2002} Tully, R.~B., Somerville, 
R.~S., Trentham, N., \& Verheijen, M.~A.~W.\ 2002, \apj, 569, 573 
\bibitem[Wong et al.(2006)]{Wong2006} Wong, O.~I., Ryan-Weber, 
E.~V., Garcia-Appadoo, D.~A., et al.\ 2006, \mnras, 371, 1855 
\bibitem[Yahil(1985)]{Yahil1985} Yahil, A.\ 1985, European 
Southern Observatory Conference and Workshop Proceedings, 20, 359 
\bibitem[Zitrin \& Brosch(2008)]{Zitrin2008} Zitrin, A., \& Brosch, N.\ 2008, \mnras, 390, 408 
\end{thebibliography}
\end{document}